\documentclass[letterpaper,twocolumn,10pt]{article}
\pdfoutput=1

\usepackage{usenix,epsfig,endnotes}
\usepackage{booktabs}
\usepackage{color}
\usepackage{epstopdf}
\usepackage[colorlinks=true,
            citecolor=blue,
            linkcolor=blue,
            urlcolor=black,
            pdfstartview=FitH,
            bookmarks=true,
            bookmarksopen=true,
            bookmarksdepth=2,
            ]{hyperref}
\usepackage[capitalize]{cleveref}
\usepackage{multirow}
\usepackage{xspace}
\usepackage{paralist}

\long\def\com#1{}

\long\def\xxx#1{ {\bf XXX:} [{\color{red} \em #1}]}

\newcommand\diagram[2]{
	\begin{figure}[t]
		\centering
		\includegraphics[width=1.00\columnwidth]
		{figures/#2}
		\caption{#1}
		\label{fig:#2}
	\end{figure}
}

\defaultleftmargin{\parindent}{}{}{}

\newcommand{\eg}{{\em e.g.}}
\newcommand{\ie}{{\em i.e.}}

\newcommand{\coinname}{ByzCoin\xspace}

\begin{document}
\date{}

\title{\Large \bf Enhancing Bitcoin Security and Performance with \\
	Strong Consistency via Collective Signing}

\author{
	Eleftherios Kokoris-Kogias,
	Philipp Jovanovic,
	Nicolas Gailly, \\
	Ismail Khoffi,
	Linus Gasser, and
	Bryan Ford \\
	EPFL}

\maketitle

\subsection*{Abstract}

While showing great promise,
Bitcoin requires users to wait tens of minutes for transactions to commit,
and even then, offering only probabilistic guarantees.
This paper introduces \coinname, a novel Byzantine consensus protocol
that leverages scalable collective signing
to commit Bitcoin transactions irreversibly within seconds.
\coinname achieves Byzantine consensus
while preserving Bitcoin's open membership
by dynamically forming hash power-proportionate consensus groups
that represent recently-successful block miners.
\coinname employs communication trees
to optimize transaction commitment and verification under normal operation 
while guaranteeing safety and liveness under Byzantine faults,
up to a near-optimal tolerance of
$f$ faulty group members among $3 f + 2$ total. 
\coinname mitigates double spending and selfish mining attacks 
by producing collectively signed transaction blocks
within one minute of transaction submission.
Tree-structured communication further reduces this latency 
to less than 30 seconds. Due to these optimizations, \coinname achieves a throughput 
higher than Paypal currently handles, with a confirmation latency of 15-20 seconds.

\section{Introduction}\label{sec:introduction}
\com{TODO fix capitalization in all the .tex, restructure sections that only
have one subsection}



Bitcoin~\cite{nakamoto08bitcoin} is a decentralized cryptocurrency
providing an open, self-regulating alternative to classic
currencies managed by central authorities such as
banks. Bitcoin builds on a peer-to-peer network where users can submit
transactions without intermediaries.
Special nodes, called \emph{miners},
collect transactions, solve computational
puzzles (\emph{proof-of-work}) to reach consensus, and add the
transactions in form of blocks to a distributed public ledger known as
the \emph{blockchain}.

The original Bitcoin paper argues that 
transaction processing is secure and irreversible,
as long as the largest colluding group of miners represents
less than $50\%$ of total computing capacity
and at least about one hour has elapsed.
This high transaction-confirmation latency
limits Bitcoin's suitability for real-time transactions.
Later work revealed additional vulnerabilities
to transaction reversibility, double-spending, and strategic mining
attacks~\cite{eyal14majority,gervais15tampering,heilman15eclipse,karame12double,nayak15stubborn,apostolaki16hijacking}.

The key problem is that Bitcoin's
consensus algorithm provides only probabilistic 
\com{I still believe that it can be called eventual~\cite{decker16bitcoin}...,,, and rachid agrees :P }%
\com{The Christian Decker arxiv preprint is not at all authoritative
	on the definition of the term "eventual consistency" -
	in fact it doesn't even try to define it properly,
	but just uses it fast-and-loose, without even referencing
	any of the classic papers on eventual consistency such as Bayou.
	\url{http://suraj.lums.edu.pk/~cs582s05/papers/06_bayou.pdf}
	In its classic definition, eventual consistency can tolerate
	arbitrarily-long delays and network partitions;
	Bitcoin consistency cannot.
	Further, eventual consistency traditionally provides
	particular consistency properties that Bitcoin does not:
	for example, see the list of properties in this CACM article
	"Eventually Consistent",
	some of which I'm pretty sure Bitcoin breaks:
	\url{http://delivery.acm.org/10.1145/1440000/1435432/p40-vogels.pdf?ip=85.244.252.161&id=1435432&acc=OPEN&key=4D4702B0C3E38B35.4D4702B0C3E38B35.4D4702B0C3E38B35.6D218144511F3437&CFID=637045753&CFTOKEN=40268510&__acm__=1467096482_35be85ff3d27a409f8f57bee22704cfc}
	I can understand Rachid thinking that any consistency model
	that is not formally guaranteed to complete in any precise time
	could be called eventual consistency, but
	that's a far more expansive use of the term than 
	what it's traditionally used for in the systems community
	where the term actually arose.  i.e., Rachid is wrong on this one.
}%
consistency guarantees.
Strong consistency could offer cryptocurrencies
three important benefits.
First, all miners instantly agree on the validity of blocks, 
without wasting computational power resolving inconsistencies
(\emph{forks}). Second, clients need not wait for extended periods
to be certain that a submitted transaction is committed; as
soon as it appears in the blockchain, the transaction can be considered
confirmed. Third, strong consistency provides \emph{forward security}:
as soon as a block has been appended to the blockchain, it stays there
forever.
Although increasing the consistency of cryptocurrencies has been suggested
before~\cite{croman16scaling,decker16bitcoin,mazieres15stellar,schwartz14ripple,vukolic15quest},
existing proposals give up Bitcoin's decentralization, and/or
introduce new and non-intuitive security assumptions,
and/or lack experimental evidence of performance and scalability.

\com{ Add to list of papers above suggesting PBFT:
	\url{http://fc16.ifca.ai/bitcoin/papers/CDE+16.pdf}.
	Also perhaps cite this paper later in Design section,
	when motivating need to make PBFT scale.  
	(See "consortium consensus" and "sharding" in sec 4.2.)

	We should probably have a paragraph somewhere later (not here)
	discussing sharding as an alternative approach,
	noting that a key problem is who decides "how to shard"
	in a trustworthy fashion (i.e., who to trust to make that decision?).
	Obvious choice is "random", but note that
	trustworthy, bias-resistant randomness is hard
	(cite Lenstra and Bonneau). }

This work introduces \coinname, a Bitcoin-like cryptocurrency
enhanced with strong consistency, based on the principles of the well-studied
Practical Byzantine Fault Tolerance (PBFT)~\cite{castro99practical}
algorithm.
\coinname addresses four key challenges in
bringing PBFT's strong consistency to cryptocurrencies:
(1)~open membership, (2)~scalability to hundreds of replicas,
(3)~proof-of-work block conflicts, and (4)~transaction commitment rate.

PBFT 
was not designed for scalability to large consensus groups:
deployments and experiments often employ the minimum of
four replicas~\cite{kotla07zyzzyva},
and generally have not explored scalability levels
beyond 7~\cite{castro99practical}
or 16 replicas~\cite{cowling06HQ,guerraoui10next,abdelmalek05QU}.
\coinname builds PBFT atop
CoSi~\cite{syta15decentralizing},
a collective signing protocol
that efficiently aggregates hundreds or thousands of signatures.
Collective signing reduces both the costs of PBFT rounds
and the costs for ``light'' clients to verify transaction commitment.
Although CoSi is not a consensus protocol,
\coinname implements Byzantine consensus
using CoSi signing rounds to make
PBFT's {\em prepare} and {\em commit} phases scalable.

PBFT normally assumes a well-defined, closed group of replicas,
conflicting with Bitcoin's open membership and use of proof-of-work
to resist Sybil attacks~\cite{douceur02sybil}. 
\coinname addresses this conflict
by forming consensus groups dynamically from {\em windows}
of recently mined blocks,
giving recent miners {\em shares} or voting power proportional
to their recent commitment of hash power.
Lastly, to reduce transaction processing latency
we adopt the idea from Bitcoin-NG~\cite{eyal16bitcoinng} 
to decouple transaction verification from block mining.

\com{	I think this is a subtle enough detail that we can leave it to
	be introduced in the design section. -baf
While strong consistency by definition does not allow any forks, 
but the natural way of solving block conflicts 
can make Selfish Mining~\cite{eyal14majority} easier. In order to avoid that 
we develop an algorithm that makes all participants 
unbiasedly decide on how to solve a fork.
}

\com{	Not sure what this text is saying or why it's important here. -baf

We also consider a slightly weaker and more realistic adversary 
so that it is possible to further enhance scalability of the system 
with tree communication patterns following the paradigm of popular multicast
protocols~\cite{castro03splitstream, venkataraman06chunkyspread}. If the adversary manages to
compromise the liveness of the highly scalable \coinname instance, 
then instead of failing, \coinname continues functioning with 
degraded performance
}

Experiments with a prototype implementation of \coinname show that 
a consensus group formed from approximately the past 24 hours 
of successful miners (144 miners)
can reach consensus in less than 20 seconds, on blocks
of Bitcoin's current maximum size (1MB). 
A larger consensus group formed from one week 
of successful miners (1008) reached consensus on an 8MB block in 
90 seconds, showing that the systems scales both with the number of 
participants and with the block size.
For the 144-participant consensus group, 
with a block size of 32MB, the system handles 974 transactions per second (TPS)
with a 68-second confirmation latency.
These experiments suggest that \coinname can handle loads
higher than PayPal and comparable with Visa.

\com{	This doesn't seem to add any experimental results. -baf
\coinname{} mitigates 0-confirmation attacks
by sending the first confirmation in less than a minute, 
eliminates other double-spending attacks
as a result of PBFT's strong consistency,
and mitigates selfish mining attacks.
}
\com{	This text sounds inconsistent with the above:
	above it says 68-second confirmation, below it says 30-second.
	If the scenarios are different and this is actually important,
	then please fix/explain/clarify. 
	Otherwise we can just drop this part. -baf

We find that \coinname enables clients to confirm transactions,
without risk of reversal or double-spending,
within 30 seconds of issuance.
To reach comparable confidence of irreversible commitment,
Bitcoin requires more than $3$ hours,
assuming as in \coinname that an attacker controls less than $25\%$ 
of hash power.
}

\coinname is still a proof-of-concept
with several limitations.
First, \coinname does not improve on Bitcoin's proof-of-work mechanism;
finding a suitable
replacement~\cite{ateniese14proofs,ford08nyms,king12ppcoin,yu08sybillimit}
is an important but orthogonal area for future work.
Like many BFT protocols in practice~\cite{clement09making,guerraoui10next}, 
\coinname is vulnerable to slowdown or temporary DoS attacks
that Byzantine nodes can trigger.
Although a malicious leader cannot violate or permanently block consensus,
he might temporarily exclude minority sets ($<\frac{1}{3}$) of victims
from the consensus process, depriving them of rewards,
and/or attempt to censor transactions.
\coinname guarantees security only against attackers who
consistently control less than a third (not 50\%) of consensus group shares --
though Bitcoin has analogous weaknesses accounting for
selfish mining~\cite{eyal14majority}.
\com{
Finally, \coinname's security is at present
analyzed only informally (\cref{sec:security}).}

\com{ From Conclusion: "Our system can be deployed to any blockchain-based system since the leader election mechanism of proof-of-work can easily change to another mechanism like proof-of-stake or be static if openness is not an issue." 

Proof-of-work is not a limitation we keep it in order to be Bitcoin-compatible -lkk}

\com{
	And another: 
	We have not yet formally analyzed the security of the protocol,
	although Section~\ref{sec:security} presents an informal analysis.
}

In this paper we make the following key contributions:

\begin{compactitem}

\item	We use collective signing~\cite{syta15decentralizing}
	to scale BFT protocols to large consensus groups
	and enable clients to verify operation commitments efficiently.

\item	We present (\S\ref{sec:arch}) 
	the first demonstrably practical Byzantine consensus protocol
	supporting not only static consensus groups
	but also dynamic membership proportional to proof-of-work as in Bitcoin.

\item	We demonstrate experimentally  (\S\ref{sec:eval})
	that a strongly-consistent cryptocurrency
	can increase Bitcoin's throughput by two orders of magnitude,
	with a transaction confirmation latency under one minute.

\com{	This claim would be appropriate if we did a serious formal analysis. :) }

\item We find through security analysis (\S\ref{sec:security})
	that \coinname can mitigate
	several known attacks on Bitcoin provided no attacker
	controls more than $\frac{1}{4}$ of hash power.

\end{compactitem} 

\com {save space, no-one reads it... important refs up in contributions

The remainder of the paper is organized as follows.
\cref{sec:back} summarizes core
concepts of Bitcoin and its variations, scalable collective signing, and
Byzantine fault tolerance.
\cref{sec:arch} details the \coinname protocol.
\cref{sec:eval} describes an evaluation of our prototype implementation
of \coinname.
\cref{sec:security} informally analyzes \coinname's security 
and \cref{sec:related} outlines related work,
}

\section{Background and Motivation}\label{sec:back}

This section first outlines the three most relevant areas
of prior work that \coinname builds on:
cryptocurrencies such as Bitcoin and Bitcoin-NG,
Byzantine fault tolerance (BFT) principles,
and collective signing techniques.

\subsection{Bitcoin and Variations}
\label{sec:bitcoin}

\paragraph{Bitcoin.}

At the core of Bitcoin~\cite{nakamoto08bitcoin} 
rests the so-called
\emph{blockchain}, a public, append-only database maintained by
\emph{miners} and serving as a global ledger of all transactions ever
issued.
Transactions are bundled into \emph{blocks}
and validated by a \emph{proof-of-work}.
A block is valid if its cryptographic hash has $d$ leading
zero bits, where the difficulty parameter $d$ is adjusted periodically
such that new blocks are mined about every ten minutes on average.
Each block includes a Merkle tree~\cite{merkle79secrecy}
of new transactions to be committed,
and a cryptographic hash chaining to the last valid block,
thereby forming the blockchain.
Upon successfully forming a new block with a valid proof-of-work,
a miner broadcasts the new block to the rest of the miners,
who (when behaving properly) accept the new block, if it extends
a valid chain strictly longer than any they have already seen.

Bitcoin's decentralized consensus and security derive from an assumption
that a majority of the miners, measured in terms of \emph{hash power}
or ability to solve hash-based proof-of-work puzzles,
follows these rules and always attempts to extend the longest existing chain.
As soon as a quorum of miners with the majority of the
network's hash power approves a given block by mining on top of it,
the block remains embedded in any future chain~\cite{garay15bitcoin}.
Bitcoin's security is guaranteed by the fact that this majority
will be extending the legitimate chain faster
than any corrupt minority that might try to rewrite history
or double-spend currency.
However, Bitcoin's consistency guarantee is only probabilistic, 
which leads to two fundamental problems.

First, multiple miners might find distinct blocks with the same parent
before the network has reached consensus. Such a conflict is called a
\emph{fork}, an inconsistency that is temporarily allowed until one
of the chains is extended yet again. Subsequently,
all well-behaved miners on the shorter chain(s) switch to the new longest one.
All transactions appearing only in the rejected block(s) are invalid and must be
resubmitted for inclusion into the winning blockchain.
This means that Bitcoin clients who want high certainty
that a transaction is complete
(\eg, that they have irrevocably received a payment)
must wait not only for the next block but for several blocks thereafter, 
thus increasing the time interval until a transaction can be considered complete.
As a rule of thumb~\cite{nakamoto08bitcoin}, a
block is considered as permanently added to the blockchain after about $6$ new
blocks have been mined on top of it,
for a confirmation latency of $60$ minutes on average.

Second, the Bitcoin block size is currently limited to $1$\,MB. 
This limitation in turn results in an upper bound on the number of
transactions per second (TPS) the Bitcoin network can handle,
estimated to be an average of $7$\,TPS. For comparison, Paypal handles
$500$\,TPS and VISA even $4000$\,TPS. An obvious solution to enlarge Bitcoin's
throughput is to increase the size of its blocks. Unfortunately,
this solution also
increases the probability of forks due to higher propagation delays and the risk
of double-spending attacks~\cite{sompolinsky13fastbitcoin,gervais16security,kiayias15speed}.
Bitcoin's liveness and security properties
depend on forks being relatively rare.
Otherwise, the miners would spend much of their effort
trying to resolve multiple forks~\cite{gervais15tampering,croman16scaling},
or in the extreme case, completely centralize Bitcoin~\cite{eyal16bitcoinng}

\com{
These issues reduce Bitcoin's suitability for applications requiring
real-time transactions and limit its widespread adoption.
The primary contribution of \coinname,
detailed in \cref{sec:arch} below,
is to address the above limitations via 
a Bitcoin-like protocol with strong consensus.
}

\paragraph{Bitcoin-NG.}\label{sec:bitcoinng}

Bitcoin-NG~\cite{eyal16bitcoinng} 
makes the important observation
that Bitcoin blocks serve two different purposes:
(1) election of a leader who decides how to resolve potential
inconsistencies, and (2) verification of transactions.
Due to this observation,
Bitcoin-NG proposes two different block types: \emph{Keyblocks} are generated
through mining with proof-of-work and are used to securely elect leaders,
at a moderate frequency, such as every 10 minutes as in Bitcoin.
\emph{Microblocks} contain transactions, require no proof-of-work, and are
generated and signed by the elected leader. 
This separation enables Bitcoin-NG to process many microblocks
between the mining of two keyblocks,
enabling transaction throughput to increase.
\com{The above sentence could/should be simplified}

Bitcoin-NG, however, retains many drawbacks
of Bitcoin's consistency model.
Temporary forks due to near-simultaneous keyblock mining,
or deliberately introduced by selfish or malicious miners,
can still throw the system into an
inconsistent state for $10$ minutes or more.
Further, within any 10-minute window
the current leader could still intentionally fork or rewrite history
and invalidate transactions. 
If a client does not wait several tens of minutes (as in Bitcoin) 
for transaction confirmation, 
he is vulnerable to double-spend attacks by the current leader
or by another miner who forks the blockchain.
Although Bitcoin-NG includes disincentives for such behavior,
these disincentives amount at most to the ``mining value'' of the keyblock
(coinbase rewards and transaction fees): 
Thus, leaders are  both able and have incentives to double-spend
on higher-value transactions.

Consequently, although Bitcoin-NG permits higher transaction throughput,
it does not solve Bitcoin's consistency weaknesses.
Nevertheless, Bitcoin-NG's decoupling
of keyblocks from microblocks is an important idea that
we build on in Section~\ref{sec:decoupling}
to support high-throughput and low-latency transactions in \coinname.



\com{
For \coinname we build on Bitcoin-NG's idea of decoupling leader-election (key blocks) 
from the transaction-processing (microblocks), 
which enables to separate the time scales of these two operations. 
However we  additionally make use
of collective signing to achieve much stronger consistency assurances than
Bitcoin-NG thereby amending many of its issues.
}

\subsection{Byzantine Fault Tolerance}
\label{sec:bft}

The \emph{Byzantine Generals' Problem}~\cite{lamport82byzantine,pease80reaching} 
refers to the situation where the malfunctioning of one or several components of
a distributed system prevents the latter from reaching an agreement. 
Pease et al.~\cite{pease80reaching} show that $3f+1$ participants are necessary 
to be able to tolerate $f$ faults and still reach consensus. The
\emph{Practical Byzantine Fault Tolerance (PBFT)}
algorithm~\cite{castro99practical} was the first efficient solution to the
Byzantine Generals' Problem that works in weakly synchronous environments
such as the Internet. PBFT offers both \emph{safety} and \emph{liveness}
provided that the above bound applies, \ie, that at most $f$ faults among
$3f+1$ participants occur. 
PBFT triggered a surge of research on Byzantine
replication algorithms with various optimizations and
trade-offs~\cite{abdelmalek05QU,cowling06HQ,kotla07zyzzyva,guerraoui10next}.
\com{	It's not clear what point the reader should learn from this blurb,
	and this is only one of many papers that critique PBFT
	and point out improvements; 
	another relevant one making a similar point is Clement et al, 
	"Making Byzantine Fault Tolerant Systems Tolerate Byzantine Faults"
	(NSDI '09).
	If any of these many follow-on papers are particularly important
	for our purposes, then state how/why. -baf
Guerraoui et al.~\cite{guerraoui10next} observed that all these protocols only optimize
the case where no Byzantine faults occur and otherwise fall back to protocols
even slower than the original PBFT. They further show how to generalize PBFT
resulting in a simpler and more efficient approach to solve the Byzantine
Generals' Problem than all previous proposals and resolving many of their
drawbacks. 
}

\com{ Briefly (in 1 paragraph) summarize the basic PBFT round operation:
	pre-prepare, prepare, commit. 
	And briefly summarize how view changes work to ensure liveness.
}

Every round of PBFT has three distinct phases. 
In the first, {\em pre-prepare} phase,
the current primary node or {\em leader} announces 
the next record that the system should agree upon. 
On receiving this pre-prepare, every node validates 
the correctness of the proposal
and multicasts a {\em prepare} message to the group. 
The nodes wait until they collect 
a quorum of ($2f+1$) prepare messages 
and publish this observation with a {\em commit} message. 
Finally, they wait for a quorum of ($2f+1$) commit messages 
to make sure that enough nodes have recorded the decision. 
\com{	I don't understand what this sentence is trying to say,
	and it doesn't sound likely to be important anyway.
	If it is, please clarify. -baf
If consequent rounds of PBFT finish out of order, 
the changes are stashed and implemented in order.
}

PBFT relies upon a correct leader to begin each round
and proceeds if a two-thirds quorum exists; 
consequently, the leader is an attack target.
For this reason PBFT has a view-change protocol 
that ensures liveness in the face of a faulty leader.
All nodes monitor the leader's actions 
and if they detect either malicious behavior or a lack of progress,
initiate a view-change. 
Each node independently announces its desire to change leaders
and stops validating the leader's actions. 
If a quorum of ($2f+1$) nodes decides that the leader is faulty, 
then the next leader in a well-known schedule takes over.

PBFT has its limitations.
First, it assumes a
fixed, well-defined group of replicas, thus
contradicting Bitcoin's basic principle of being
decentralized and open for anyone to participate.
Second, each PBFT replica normally communicates directly
with every other replica during each consensus round,
resulting in $O(n^2)$ communication complexity:
This is acceptable when $n$ is typically 4 or not much more,
but becomes impractical if $n$ represents
hundreds or thousands of Bitcoin nodes.
Third, after submitting a transaction to a PBFT service,
a client must communicate with a super-majority of the replicas
in order to confirm the transaction has been committed and to learn its outcome,
making secure transaction {\em verification} unscalable.




\subsection{Scalable Collective Signing}\label{sec:CoSi}


CoSi~\cite{syta15decentralizing}
is a protocol for scalable collective signing,
which enables an authority or \emph{leader}
to request that statements be publicly validated and
\emph{(co-)signed} by a decentralized group of \emph{witnesses}.
Each protocol run yields a {\em collective signature}
having size and verification cost comparable to an individual signature,
but which compactly attests that both the leader and its
(perhaps many) witnesses observed and agreed to sign the statement.

\begin{figure}[t]
    \centering
    \includegraphics[width=0.90\columnwidth]{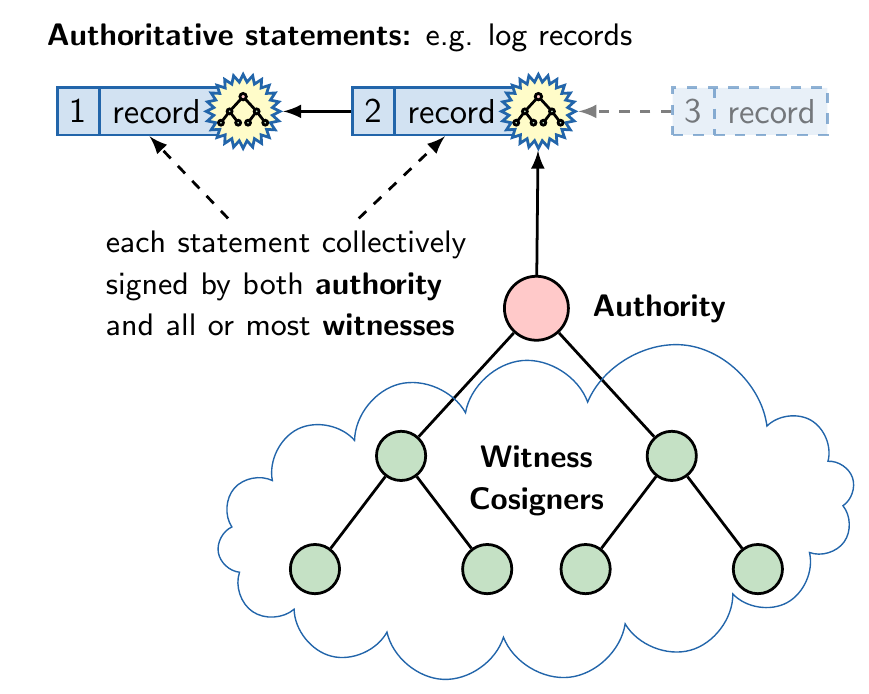}
    \caption{CoSi protocol architecture}
    \label{fig:cosi}
\end{figure}

\com{	Not sure that this adds anything the above paragraph doesn't say? -baf
On a high level the CoSi protocol works as follows: the designated leader
implements the authoritative logic and proposes statements to be signed in a
round-based fashion. The roster of witnesses verify the leader's
signing-proposals according to application-specific rules and if found to be
valid co-sign the message. An overview of the protocol architecture in the
context of a tamper-evident logging mechanism is given in~\cref{fig:cosi}.
}

To achieve scalability,
CoSi combines Schnorr multi-signatures~\cite{schnorr91efficient}
with communication trees that are long used in multicast
protocols~\cite{castro03splitstream,deering90multicast,venkataraman06chunkyspread}.
\com{	Why do we need this level of low-level detail?
Let $\mathcal{G}$ be a group of prime order $q$ in which the discrete logarithm
problem is believed to be hard. For practical purposes, $\mathcal{G}$ is usually
an elliptic curve. Further let $G \in \mathcal{G}$ be a generator of the group.
We assume that  there are $n$ nodes which participate in the CoSi protocol
indexed by $0 \leq i \leq n-1$. Among those nodes we identify w.l.o.g. by $i =
0$ the leader and by $1 \leq i \leq n-1$ the witnesses. We assume
that each node has chosen a random value $x_i < q$ as its \emph{private key} and
has computed the corresponding \emph{public key} via $X_i = G^{x_i}$. Moreover,
we assume that the communication tree has already been setup and that each node
knows which parent and children it has (if any). For further information on that
matter we refer to the original publication~\cite{syta15decentralizing}.
}
Initially, the protocol assumes that signature verifiers know the public keys
of the leader and those of its witnesses,
all of which combine to form a well-known aggregate public key.
For each message to be collectively signed,
the leader then initiates a CoSi four-phase protocol round
that require two round-trips over the communication tree
between the leader and its witnesses:

\begin{compactenum}

    \item \textbf{Announcement:} The leader broadcasts an announcement of a new
        round down the communication tree. The announcement can optionally
        include the message $M$ to be signed, otherwise $M$ is sent in
        phase three.

    \item \textbf{Commitment:} Each node picks
        a random secret and uses it to compute a Schnorr commitment.
        In a bottom-up process, each node obtains an aggregate
	Schnorr commitment from its immediate children,
	combines those with its own commitment,
	and passes a further-aggregated commitment up the tree.

    \item \textbf{Challenge:} The leader computes
	a collective Schnorr challenge using a cryptographic hash function
        and broadcasts it down the communication tree,
	along with the message $M$ to sign, if the latter has not
        already been sent in phase one.

    \item \textbf{Response:} Using the collective challenge, all nodes
        compute an aggregate response in a bottom-up fashion that
	mirrors the commitment phase.

\end{compactenum}

The result of this four-phase protocol is the production of
a standard Schnorr signature that
requires about 64 bytes,
using the Ed25519 elliptic curve~\cite{bernstein12highspeed}, and
that anyone can verify against the aggregate public key
nearly as efficiently as the verification of an individual signature.
Practical caveats apply if some witnesses are offline
during the collective signing process:
in this case the CoSi protocol can proceed,
but the resulting signature grows to include metadata
verifiably documenting which witnesses did and did not co-sign.
We refer to the CoSi paper for details~\cite{syta15decentralizing}.

\section{ByzCoin Design}\label{sec:arch}

This section presents \coinname
with a step-by-step approach,
starting from a simple ``strawman'' combination
of PBFT and Bitcoin.
From this strawman, we progressively address the challenges of
determining consensus group membership,
adapting Bitcoin incentives and mining rewards,
making the PBFT protocol scale to large groups
and handling block conflicts and selfish mining.

\subsection{System Model}

\coinname is designed for untrustworthy networks
that can arbitrarily delay, drop, re-order or duplicate messages. 
To avoid the FLP impossibility~\cite{fischer85impossibility},
we assume the network has a weak synchrony property~\cite{castro99practical}.
The \coinname system is comprised of a set of $N$  block miners 
that can
generate key-pairs, but there is no trusted public-key infrastructure.
Each node $i$ has a limited amount of 
\emph{hash power} that corresponds to the maximum number
of block-header hashes the node can perform per second.

At any time $t$ a subset of miners $M\left(t\right)$ is controlled by a 
malicious attacker and are considered faulty.
Byzantine miners can behave arbitrarily, diverting from the protocol
and colluding to attack the system.
The remaining honest miners follow the prescribed protocol.
We assume that the total hash power of all Byzantine nodes
is less than $\frac{1}{4}$ of the system's total hash power at any time, since
proof-of-work-based cryptocurrencies become vulnerable to selfish mining attacks
by stronger adversaries~\cite{eyal14majority}.
 
\subsection{Strawman Design: PBFTCoin}

For simplicity, we begin with PBFTCoin,
an unrealistically simple protocol that naively combines PBFT with Bitcoin,
then gradually refine it into \coinname.

For now, we simply assume that a group of $n =
3f + 1$ PBFT replicas, which we call \emph{trustees},
has been fixed and globally agreed upon upfront,
and that at most $f$ of these trustees are faulty.
As in PBFT, at any given time, one of these trustees is the {\em leader},
who proposes transactions and drives the consensus process.
These trustees collectively maintain a Bitcoin-like blockchain,
collecting transactions from clients and appending them via new blocks,
while guaranteeing that only one blockchain history ever exists
and that it can never be rolled back or rewritten.
Prior work has suggested essentially such a
design~\cite{croman16scaling,decker16bitcoin},
though without addressing the scalability challenges it creates.

\com{	Not necessarily true, if proof-of-work is still the way
	fresh currency is brought into existence!
Since this group is generally trusted, there is no
need to have miners or a proof-of-work mechanism.
}
\com{	Not sure what useful detail this adds.
A client that wants to submit
a transaction, sends the latter to the group of trustees which in turn run one
iteration of PBFTCoin and, after the blockchain has been updated successfully,
return a notification to the client confirming its transaction.
}

Under these simplifying assumptions, PBFTCoin guarantees safety and liveness,
as at most $f$
nodes are faulty and thus the usual BFT security bounds apply. However, the
assumption of a fixed group of trustees is 
unrealistic for Bitcoin-like decentralized cryptocurrencies
that permit open membership.
Moreover, as PBFT trustees authenticate each other
via non-transferable symmetric-key MACs,
each trustee must communicate directly with most other trustees in every round,
thus yielding $O(n^2)$ communication complexity.

Subsequent sections address these restrictions,
transforming PBFTCoin into \coinname in four main steps:

\begin{compactenum}

\item We use Bitcoin's proof-of-work mechanism to determine
	consensus groups dynamically
	while preserving Bitcoin's defense against Sybil attacks.

\item We replace MAC-authenticated direct communication
	with digital signatures to make authentication transferable
	and thereby enabling sparser communication patterns that can 
	reduce the normal case communication latency
	from $O(n^2)$ to $O(n)$.

\item We employ scalable collective signing to reduce per-round
	communication complexity further to $O(\log n)$ and
	reduce typical signature verification complexity
	from $O(n)$ to $O(1)$. 

\item We decouple transaction verification from leader election to
    achieve a higher transaction throughput.

\end{compactenum}


\subsection{Opening the Consensus Group}
\label{sec:membership}


\com{
\xxx{ I think we need a better term for "virtual miners".
	I'm thinking maybe "mining shares" or just "shares"?
	Mining a key block creates one share, whose lifetime is
	one group-size "window" of keyblocks.
	At any given moment (within any given window),
	each miner has a number of "consensus group votes" or "memberships"
	corresponding to the number of shares they mined within that window.
	In other words, I think shares (or a similar term)
	makes it more obvious that these are purely virtual entities,
	and that it is OK and normal for miners to hold multiple shares
	at a given time. }

\xxx{ Replaced ``virtual miners'' by ``consensus group share'' -- pj}
}

Removing PBFTCoin's assumption of
a closed consensus group of trustees
presents two conflicting challenges.
On the one hand,
conventional BFT schemes rely on a well-defined consensus group
to guarantee safety and liveness.
On the other hand,
Sybil attacks~\cite{douceur02sybil} can trivially break
any open-membership protocol involving security thresholds,
such as PBFT's assumption that at most $f$ out of $3f+1$ members are honest.

Bitcoin and many of its variations employ a mechanism already suited
to this problem: proof-of-work mining.
Only miners who have dedicated resources are
allowed to become a member of the consensus group. In refining PBFTCoin,
we adapt Bitcoin's proof-of-work mining into a
\emph{proof-of-membership} mechanism.
This mechanism maintains the ``balance of power'' within
the BFT consensus group over a given fixed-size sliding \emph{share window}. 
Each time a miner finds a new block, it receives a
\emph{consensus group share},
which proves the miner's membership in the group of
trustees and moves the share window one step forward.
Old shares beyond the current window expire and become useless
for purposes of consensus group membership.
Miners holding no more valid shares in the current window lose their membership in the consensus
group, hence they are no longer allowed to participate in the decision-making.

At a given moment in time,
each miner wields ``voting power'' of a number of shares
equal to the number of blocks the miner has successfully mined
within the current window.
Assuming collective hash power is relatively stable,
this implies that within a window,
each active miner wields a number of shares statistically proportionate to
the amount of hash power that the miner has contributed during this time period.

\com{
\xxx{ Not sure epoch is the right term here; to me epochs sound like
	well-defined time boundaries and not a sliding window.
	Perhaps just "windows"? or "group windows"? }
\xxx{ Agreed. Removed it. Maybe using ``share window'' to point out the relation to the shares? -- PJ }
}

\begin{figure}[t]
     \centering
     \includegraphics[width=1.00\columnwidth]{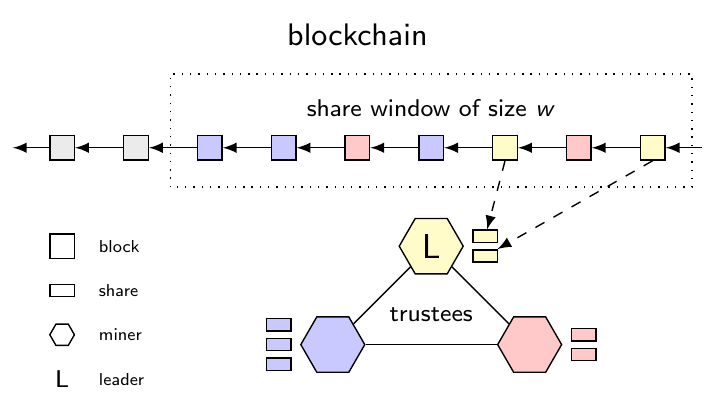}
     \caption{Valid shares for mined blocks in the blockchain are credited to miners}
     \label{fig:byzcoin-miners}
 \end{figure}

The size $w$ of the share window is defined by the average block-mining rate
over a given time frame and influences certain properties
such as the resilience of
the protocol to faults. 
For example, if we assume an average block-mining rate of $10$ minutes and set
the duration of the time frame to one day (or one week), then 
$w = 144$ ($w = 1008$). This mechanism limits the membership of miners to recently
active ones,
which prevents the system from becoming unavailable due to too many
trustees becoming inactive over time,
or from miners aggregating many shares over an
extended period and threatening the balance in the consensus group.
The relationship between blocks, miners and shares is illustrated
in~\cref{fig:byzcoin-miners}.

\com{
\xxx{ Minor nitpick about fig 2: the shares and nodes aren't labeled.
	Perhaps just add a text label "node" to the left of the triangle
	with light dotted arrows going from that to the red and blue circles,
	and add a text label "shares" to the bottom-right just under
	the little blue and/or yellow boxes.
	At any rate, just try to make the figure visually self-explanatory.}
}
\com{addressed in security section maybe need to move some parts there
We leave to future work a detailed exploration of
how large the window should be in practice.
On the one hand,
shorter windows produce smaller consensus groups, that
mitigate BFT scalability challenges,
but this produces coarser ``samples'' of recent hash power over a shorter period,
potentially leaving the system vulnerable to ``flash'' attackers who might 
invest a great deal of hash power over a short time period.
On the other hand, longer windows
produce more accurate samples of hash power over longer periods,
but could create an increased risk to the system's liveness
if there is too much ``churn'' in hash power over the course of the window:
\eg, if miners representing more than one-third of in-window blocks
suddenly leave the network before their shares have expired.

}

\paragraph{Mining Rewards and Transaction Fees.}
As we can no longer assume voluntary participation as in PBFTCoin's closed group of trustees,
we need an incentive for nodes to obtain shares in the consensus group
and to remain active.
For this purpose, we adopt Bitcoin's existing incentives
of mining rewards and transaction fees.
But instead of these rewards all going to the miner of the most recent block
we split a new block's rewards and fees across
all members of the current consensus group,
in proportion to the number of shares each miner holds.
As a consequence,
the more hash power a miner has devoted within the current window,
hence the more shares the miner holds,
the more revenue the miner
receives during payouts in the current window.
This division of rewards also creates incentives for consensus group members
to remain live and participate,
because they receive their share of the rewards for new blocks
only if they continually participate,
in particular contributing to the prepare and commit phases
of each BFT consensus round.
\com{ moved in security analysis
One problem is that the leader might censor transactions,
as in our current protocol the leader controls
which blocks are proposed and the set of transactions they contain.
Although some similar censorship risks already exist in Bitcoin
and we defer a full solution to future work,
we mitigate this risk in part via PBFT's standard view-change protocol.
Any node which detects
that the leader is dishonestly ignoring or omitting transactions
from live nodes, holds no more shares,
or has failed in other detectable ways,
broadcasts a view-change message and stops accepting blocks from the current
leader. When $2f+1$ nodes send a view-change message,
the previous leader loses authority on processing further transactions
and the next leader takes over.

\com{Include ``transaction proposal" in the commit phase of prepare round so that
censorship is a lot harder. If a node does not see the transactions he proposes in the final proposed microblock he refuses to sign and notifies his buddies of the problem. If the leader insists on his decision the system will soon halt. Of course this attack is only possible IF someone gets  more than 1/3 which is very unlikely in the new threat model. }

A second limitation of the current protocol
is that a malicious leader might choose
to send a message to only slightly over $\frac{2}{3}$
of current consensus group members and still make progress,
depriving rewards from up to $\frac{1}{3}$ of members
who might indeed be online but ``out of favor'' with the leader.
Whereas we again defer a full solution to future work,
our current design at least removes financial incentives
for such strategies
by specifying that the reward-portions of purportedly ``offline'' trustees 
are not redistributed to those online but are merely discarded.
As with the transaction censorship problem,
honest nodes could also, in principle, gossip with each other
to detect a leader's malicious inclusion of live members
and in response initiate a view-change.

}

\com{Maybe explain e bit more why the shares exist? Because we want to maintain the proof-of-work purpose? See commented example below -lkk}

\subsection{Replacing MACs by Digital Signatures}

In our next refinement step towards \coinname,
we tackle the scalability challenge resulting from
PBFT's \com{this is actually not completely true since in PBFT over a LAN the
consensus is done using IP multicast over UDP and the MACS are computed in arrays to provide third party verifiability. However, since multicast is hard to implement on the internet, implementing PBFT on Bitcoin would yield $O(n^2)$

non-transferable MAC-based message authentication.
As MACs rely on pairwise shared secrets, to reach consensus
all trustees must exchange 
messages pairwise in each round, thus
resulting in a}%
typical communication complexity of $O(n^2)$, where $n$ is the group size.
PBFT's choice of MAC-authenticated all-to-all communication
was motivated by the
desire to avoid public-key operations on the critical transaction path. However,
the cost for public-key operations has decreased due
to well-optimized asymmetric cryptosystems~\cite{bernstein12highspeed}, 
making those costs less of an issue.

By adopting digital signatures for authentication, we are able to use sparser
and more scalable communication topologies, thus enabling the current leader
to collect and distribute third-party verifiable evidence that certain steps in
PBFT have succeeded. This removes the necessity for all trustees to communicate
directly with each other. With this measure 
we can either enable the leader
to collect and distribute the digital signatures,
or let nodes communicate in a chain~\cite{guerraoui10next},
reducing the normal-case number of messages from $O(n^2)$ to $O(n)$.

\com{
\xxx{ This subsection
should explain how use of signatures (in general)
allows physical communication patterns to be made less dense and more scalable
by allowing the current leader (or view-owner) to collect and distribute
third-party-verifiable evidence that a Prepare or a Commit has succeeded.
An illustration would probably help here.

Also, important to note how PBFT's choice of direct communication was motivated
by the desire to avoid public-key operations on the critical transaction path,
but since then, (a) the cost of well-optimized
public key operations on elliptic curves such as Ed25519~\cite{bernstein12highspeed}
have drastically decreased, making this cost less of a critical issue; and
(b) the benefit of third-party-verifiability of signatures makes
the scalability benefit of sparser communication topologies outweigh
the absolute cost of the digital signatures when $N$ is large. 
}
}

\subsection{Scalable Collective Signing}

Even with signatures providing transferable authentication,
the need for the leader to collect and distribute --
and for all nodes to verify --
many individual signatures per round
can still present a scalability bottleneck. 
Distributing and verifying tens or even a hundred individual signatures
per round might be practical.
If we want consensus groups with a thousand or more nodes, however
(\eg, representing all blocks successfully mined in the past week),
it is costly for the leader to distribute $1000$ digital signatures
and wait for everyone to verify them. 
To tackle this challenge, we build on
the CoSi protocol~\cite{syta15decentralizing}
for collective signing.
CoSi does not directly implement consensus or BFT,
but it offers a primitive that the leader in a BFT protocol can use
to collect and aggregate prepare and commit messages during PBFT rounds.

We implement a single \coinname round by using two sequential CoSi rounds
initiated by the current leader (\ie, the owner of the current view).
The leader's announcement of the first CoSi round
(phase 1 in~\cref{sec:CoSi})
implements the {\em pre-prepare} phase
in the standard PBFT protocol (\cref{sec:bft}).
The collective signature resulting from this first CoSi round
implements the PBFT protocol's {\em prepare} phase,
in which the leader obtains attestations
from a two-thirds super-majority quorum of consensus group members
that the leader's proposal is safe and consistent with
all previously-committed history.

As in PBFT, this prepare phase ensures
that a proposal {\em can be} committed consistently,
but by itself it is insufficient to ensure
that the proposal {\em will be} committed.
The leader and/or some number of other members could fail
before a super-majority of nodes learn about the successful prepare phase.
The \coinname leader therefore initiates a second CoSi round
to implement the PBFT protocol's {\em commit} phase,
in which the leader obtains attestations from a two-thirds super-majority
that all the signing members witnessed the successful result
of the prepare phase and made a positive commitment to remember the decision.
This collective signature, resulting from this second CoSi round,
effectively attests that a two-thirds super-majority of members
not only considers the leader's proposal ``safe'' but promises to remember it,
indicating that the leader's proposal is fully committed.

\com{ A high-level diagram might be useful here, though perhaps not essential.}

In cryptocurrency terms,
the collective signature resulting from the prepare phase
provides a proof-of-acceptance of a proposed block of transactions.
This block is not yet committed, however,
since a Byzantine leader that does not publish the accepted block 
could double-spend by proposing a conflicting block in the next round.
In the second CoSi commit round,
the leader announces the proof-of-acceptance to all members,
who then validate it 
and collectively sign the block's hash
to produce a collective commit signature on the block.
This way a Byzantine leader cannot rewrite history or double-spend,
because by counting arguments at least one honest node would have
to sign the commit phase of both histories,
which an honest node by definition would not do.


The use of CoSi does not affect the fundamental principles or semantics of PBFT
but improves its scalability and efficiency in two main ways.
First, during the commit round where each consensus group member
must verify that a super-majority of members have signed the prior prepare phase,
each participant generally needs to receive only an $O(1)$-size
rather than $O(n)$-size message,
and to expend only $O(1)$ rather than $O(n)$ computation effort
by verifying a single collective signature instead of $n$ individual ones.
This benefit directly increases the scalability and reduces
the bandwidth and computation costs of consensus rounds themselves.

A second benefit
is that after the final CoSi commit round has completed,
the final resulting collective commit signature
serves as a typically $O(1)$-size proof,
which anyone can verify in $O(1)$ computation time
that a given block -- hence any transaction within that block --
has been irreversibly committed.
This secondary scalability-benefit might in practice be more important than the first,
because it enables ``light clients'' who neither mine blocks
nor store the entire blockchain history
to verify quickly and efficiently that a transaction has committed,
without requiring active communication with or having to trust
any particular full node.

\subsection{Decoupling Transaction Verification from Leader Election}
\label{sec:decoupling}

Although \coinname so far provides a scalable guarantee of strong consistency,
thus ensuring that clients need to wait only for the next block
rather than the next several blocks to verify that a transaction has committed,
the time they still have to wait {\em between} blocks can, nevertheless, be significant:
\eg, up to 10 minutes using Bitcoin's difficulty tuning scheme.
Whereas \coinname's strong consistency might in principle
make it ``safe'' from a consistency perspective to increase block mining rate,
doing so could still exacerbate liveness and other performance issues,
as in Bitcoin~\cite{nakamoto08bitcoin}.
To enable lower client-perceived transaction latency,
therefore, we build on 
the idea of Bitcoin-NG~\cite{eyal16bitcoinng}
to decouple the functions of transaction verification from block mining
for leader election and consensus group membership.

As in Bitcoin-NG, we use two different kinds of blocks.
The first, {\em microblocks} or transaction blocks,
represent transactions to be stored and committed.
The current leader creates a new microblock
every few seconds, depending on the size of the block,
and uses the CoSi-based PBFT protocol above to commit and collectively sign it.
The other type of block, {\em keyblocks},
are mined via proof-of-work as in Bitcoin
and serve to elect leaders and create shares
in \coinname's group membership protocol
as discussed earlier in Section~\ref{sec:membership}.
As in Bitcoin-NG,
this decoupling allows the current leader to propose and commit
many microblocks that contain many smaller batches of transactions,
within one $\approx 10$-minute keyblock mining period.
Unlike Bitcoin-NG,
in which a malicious leader could rewrite history or double-spend
within this period until the next keyblock,
\coinname ensures that each microblock is irreversibly committed
regardless of the current leader's behavior.

\begin{figure}[t]
    \centering
    \includegraphics[width=0.95\columnwidth]{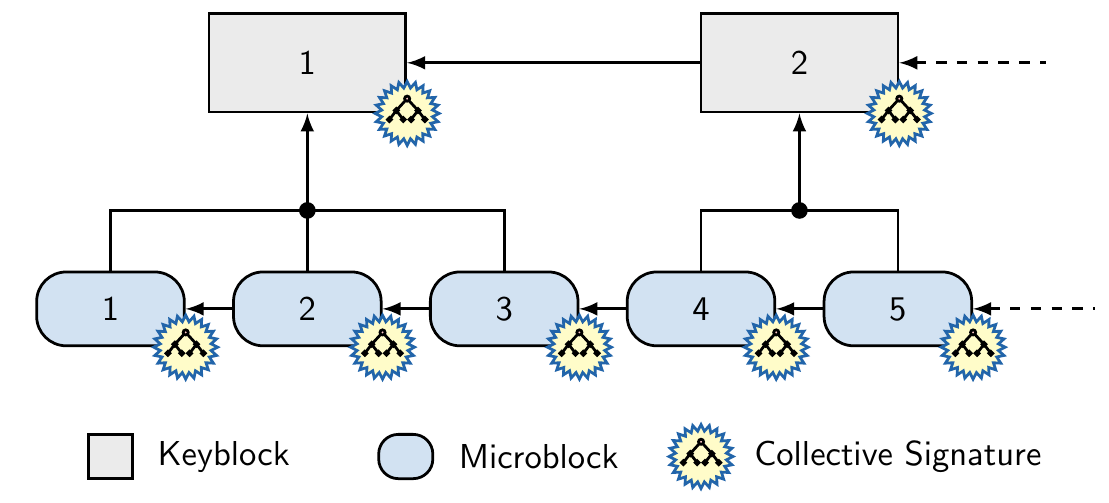}
    \caption{\coinname blockchain: Two parallel chains store information about
    the leaders (keyblocks) and the transactions (microblocks)}
    \label{fig:byzcoin-blocks}
\end{figure}

In Bitcoin-NG one blockchain includes both types of blocks,
which introduces a race condition for miners.
As microblocks are created, 
the miners have to change the header of their keyblocks to mine on top of the latest microblock.
\com{	I'm not sure what important point this text is trying to make,
	and it sounds like it's speculating on likely mining strategies,
	which is problematic and tricky to argue about.
	Perhaps better just to focus on what ByzCoin does? -baf
If they do this they must discard the work done so far and restart,.
We believe that their strategy will change 
so that they start mining on top of one microblock 
and wait until enough transactional fees 
have been added inside the microblocks,
before starting to mine on top of them. This practice will lead to 
a big number of microblocks invalidated every time a new keyblock is mined 
and if the fees are not enough (like now in Bitcoin), 
the miners will mine keyblocks to get the reward
and invalidate all microblocks, breaking the purpose of the system.
}%
In \coinname, in contrast, the blockchain becomes 
two separate parallel blockchains, as shown in~\cref{fig:byzcoin-blocks}. 
The main blockchain is the keyblock chain,
consisting of all mined blocks. 
The microblock chain is a secondary blockchain 
that depends on the primary to identify the era in 
which every microblock belongs to, 
\ie, which miners are authoritative to sign it and who is the leader of the era.

\paragraph{Microblocks.}

A microblock is a simple block that the current consensus group produces
every few seconds to represent newly-committed transactions.
Each microblock includes a set of transactions
and a collective signature. 
Each microblock also includes hashes referring to
the previous microblock and keyblock:
the former to ensure total ordering,
and the latter indicating
which consensus group window and leader created the microblock's signature.
The microblock's hash is collectively signed
by the corresponding consensus group.

\paragraph{Keyblocks.}\label{sec:keyblock}

Each keyblock contains a proof-of-work,
which is used to determine consensus group membership
via the sliding-window mechanism discussed earlier,
and to pay signers their rewards.
\com{	??? I don't understand this. -baf
as adding $N$ transactions inside every microblock 
would cause bigger blocks and slower propagation. 
In order to know the keyblock contents before the mining process begins,
the miner should include the fee payments of some round before the previous one.
}
Each newly-mined keyblock defines a new consensus group,
and hence a new set of public keys with which
the next era's microblocks will be collectively signed.
Since each successive consensus group differs from the last
in at most one member,
PBFT ensures the microblock chain's consistency and continuity
across this group membership change
provided at most $f$ out of $3f+2$ members are faulty.

Bitcoin-NG relies on incentives to discourage the next leader
from accidentally or maliciously ``forgetting''
a prior leader's microblocks.
In contrast, the honest super-majority in a \coinname consensus group 
will refuse to allow a malicious or amnesiac leader to extend
any but the most recently-committed microblock,
regardless of which (current or previous) consensus group committed it.
\com{As mentioned in the e-mail, I feel this implies that we allow forks, which is not the case, as we do not CoSi the blocks.}
Thus,
although competing keyblock conflicts may still appear,
these ``forks" cannot yield an inconsistent microblock chain.
Instead, a keyblock conflict can at worst
temporarily interrupt the PBFT protocol's liveness,
until it is resolved as mentioned in~\cref{sec:conflicts}.
\com{if miners ``follow'' the two competing keyblocks in nearly equal numbers,
such that the consensus group defined by each keyblock 
contains too few active members to commit new microblocks.
Such a (hopefully rare) interruption will be healed
as soon as one of the miners collect competing keyblock is further extended
\xxx{This is also said lower, but I really feel that overruling f+1 commits should not happen, as it does not happen in PBFT (once a client gets f+1 acks from the replicas he is good to go)}
, just as in Bitcoin.
All honest miners immediately jump to the longest keyblock chain,
and thereby re-establish liveness within the consensus group
that this new ``tie-breaking'' keyblock defines.}

\begin{figure}[t]
    \centering
    \includegraphics[width=1.00\columnwidth]{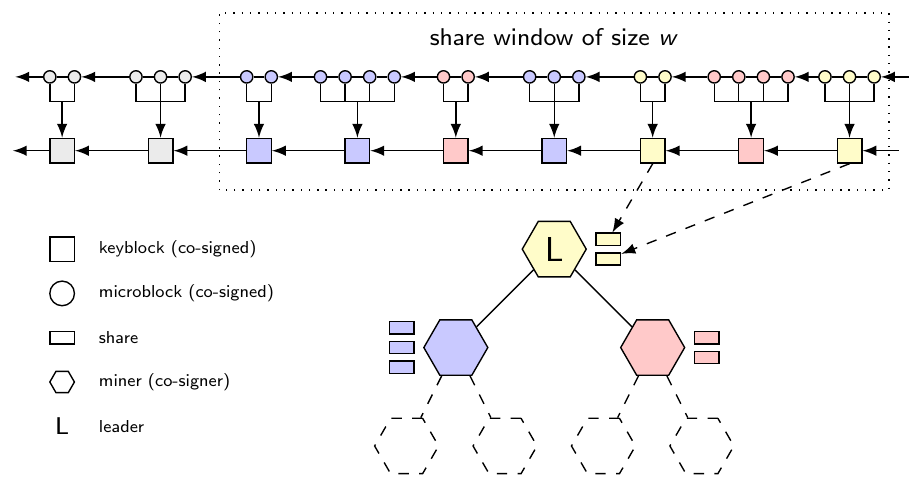}
    \caption{Overview of the full \coinname design}
    \label{fig:byzcoin-complete}
\end{figure}

Decoupling transacton verification from leader election
and consensus group evolution in this way
brings the overall \coinname architecture to completion,
as illustrated in 
\cref{fig:byzcoin-complete}.
Subsequent sections discuss further implications and challenges
this architecture presents.

\subsubsection{Keyblock Conflicts and Selfish Mining}\label{sec:conflicts}

PBFT's strong consistency by definition does not permit inconsistencies
such as forks in the microblock chain.
The way the miners collectively decide how to resolve keyblock conflicts,
however,
can still allow selfish mining~\cite{eyal14majority} to occur
as in Bitcoin. 
Worse, if the miners decide randomly to follow one of the two blocks,
then keyblock forks might frequently lead to PBFT liveness interruptions
as discussed above,
by splitting miners ``too evenly'' between competing keyblocks.
Another approach to deciding between competing keyblocks
is to impose a deterministic priority function on their hash values,
such as ``smallest hash wins.''
Unfortunately, this practice can encourage selfish mining. \com{as explained in 
Appendix A.}

One way to break a tie without helping selfish miners,
is to increase the entropy of 
the output of the deterministic prioritization function. 
We implement this idea using the following algorithm.
When a miner detects a keyblock fork,
it places all competing blocks' header hashes into a sorted array,
from low to high hash values.
The array itself is then hashed,
and the final bit(s) of this hash determine 
which keyblock wins the fork. 


\begin{figure}[t]
    \centering
    \includegraphics[width=0.65\columnwidth]{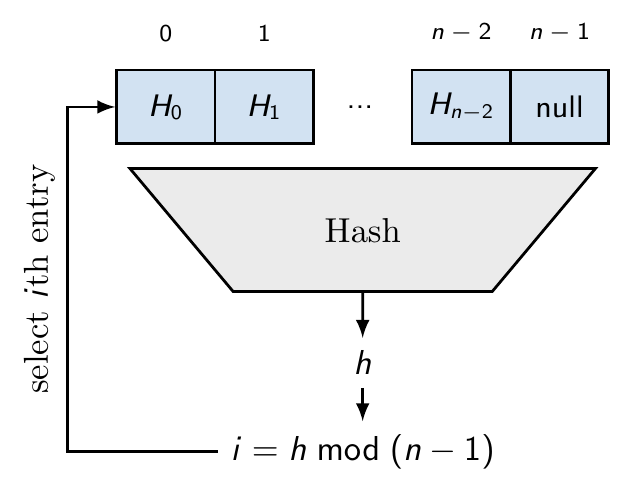}
    \caption{Deterministic fork resolution in \coinname}
    \label{fig:forks}
\end{figure}

This solution, shown in \cref{fig:forks}, also uses the idea of a 
deterministic function applied to the blocks, thus requiring no voting.
Its advantage is that the input of the hash function is partially unknown 
before the fork occurs, thus the entropy of the output is high
and difficult for 
an attacker to be able to optimize. 
Given that the search space for a possible conflict is as big as the search 
space for a new block, trying to decide if a block has better than $50\%$ 
probability of winning the fork is as hard as finding a new block.

\subsubsection{Leader Election and PBFT View Changes}

Decoupling transaction verification from the block-mining process comes at a cost. 
So far we have assumed, as in PBFT,
that the leader remains fixed unless he fails. 
If we keep this assumption, then this leader gains the power of deciding 
which transactions are verified,
hence we forfeit the fairness-enforcing benefit of Bitcoin's leader election.
To resolve this issue, every time 
a keyblock is signed, \coinname forces a mandatory PBFT view-change
to the keyblock's miner. 
This way the power of verifying transactions in blocks is 
assigned to the rightful miner, 
who has an era of microblock creation from the moment his keyblock is signed 
until the next keyblock is signed.

When a keyblock conflict occurs, 
more than one such ``mandatory" view-change occurs,
with the successful miners trying to convince other participants
to adopt their keyblock and its associated consensus group.
For example, in a keyblock fork,
one of the two competing keyblocks wins the resolution algorithm
described above.
However, if the miner of the ``losing'' block races
to broadcast its keyblock and 
more than 33\% honest miners have already committed to it
before learning about the competing keyblock,
then the ``winning" miner 
is too late and the system either
commits to the first block
or (in the worst case) loses liveness temporarily as discussed above.
This occurs because already-committed miners
will not accept a mandatory view-change except to a keyblock
that represents their committed state \com{strictly more hash power 
here it implies that f+1 a commit might be overruled, if a "losing" chain becomes longer, this is how I would manage to selfish mine.}  
and whose microblock chain extends all previously-committed microblocks.
Further analysis of how linearizability is preserved across view-changes
may be found in the original PBFT paper~\cite{castro99practical}.

\subsubsection{Tree Creation in \coinname} \label{sec:leader}

\com{ Shouldn't tree creation be part of the CoSi subsection?
	Basically, mention that CoSi operates most efficiently
	on a communication tree, but we need to build these trees,
	and here's how?
	Perhaps some of this is specific to the Bitcoin-NG design,
	but whatever part isn't specific to the decoupled blockchains
	should probably be summarized and moved up into the CoSi subsection.

	On second thought, perhaps too big of a structure change at this point,
	but may want to revisit this later. -baf
}

\com{ Up until this point the leader is static or slowly changing like CoSi, here we introduce how we solve the fact that the leader changes almost every round, without suffering the tree-creation overhead}

\com{ I don't understand this: what is "each round" below?  Each keyblock, or each microblock? }

Once a miner successfully creates a new keyblock,
he needs to form a CoSi communication tree for collective signing,
with himself as the leader.
If all miners individually acknowledge this keyblock
to transition to the next view, 
this coordination normally requires $ O\left(N\right)$ messages. 
To avoid this overhead at the beginning of each keyblock round,
the miners autonomously create the next round's tree bottom-up
during the previous keyblock round. 
This can be done in $O\left( 1 \right)$ by using the blockchain as an 
array that represents a full tree.

This tree-building process has three useful side-effects.
First, the previous leader is the first to get the new block, hence he stops creating microblocks 
and wasting resources by trying to sign them.
\com{Second, as powerful miners control multiple nodes in the tree, 
when one node gets the new block announcement, 
all of the miner's virtual miner nodes could propagate it in different levels of the tree 
which can lead to quicker spread of the news.}
Second, in the case of a keyblock conflict,
potential leaders use the same tree, and the propagation pattern is the same;
this means that all nodes will learn and decide on the conflict quickly.
Finally, in the case of a view change, the new view will be the last view that worked correctly. 
So if the leader of the keyblock $i$ fails, the next leader will again be the miner of keyblock $i-1$.

\com{ What exactly is this figure supposed to illustrate,
	in this context?  This is an extremely weird place
	for an "overview" figure to appear!!! 
	Move it somewhere appropriate, or just delete, if it's not serving a 
	critical purpose?}
\com{ I think we should definitely keep this figure since it shows the complete
\coinname design. I agree that this particular subsection is probably not the
perfect place for it. However, at the end of Section 3.6 we finally have all the
pieces together to describe the full \coinname. Another option might be to have
it as a prospect at the end of Section 3.2 but that seems to be also a bit weird
for other reasons.}
\com{Maybe before 3.6.1? -lef}

\subsection{Tolerating Churn and Byzantine Faults}
In this section we discuss the challenges of fault tolerance in \coinname,
particularly tree failures
and maximum tolerance for Byzantine faults.

\subsubsection{Tree Fault Tolerance}\label{sec:FT}


In CoSi, there are multiple different mechanisms that
allow substantial fault-tolerance. 
Furthermore the strict membership requirements 
and the frequent tree changes of \coinname{} 
increase the difficulty for a malicious attacker with less than around 25\% 
of the total hash power to compromise the system. 
A security analysis, however, must assume that a Byzantine adversary 
is able to get the correct nodes of the \coinname{} signing tree 
so that it can compromise the liveness of the system by a simple DoS.

To mitigate this risk,
we focus on recent Byzantine fault tolerance results~\cite{guerraoui10next},
modifying \coinname{} 
so that the tree communication pattern is
a normal-case performance optimization
that can withstand most malicious attacks. 
But when the liveness of the tree-based \coinname{} is compromised, 
the leader can return to non-tree-based communication 
until the end of his era.

The leader detects that the tree has failed with the following mechanism:
After sending the block to his children, the leader starts a timer 
that expires before the view-change timer. Then he broadcasts
the hash of the block he proposed and waits. 
When the nodes receive this message they check if they have seen the block
and either send an ACK or wait until they see the block and then send the ACK.
The leader collects and counts the ACKs, to detect if his block is rejected
simply because it never reaches the witnesses. 
If the timer expires or a block rejection arrives
before he receives two-thirds of the ACKs, 
the leader knows that the tree has failed and reverts
to a flat \coinname{} structure
before the witnesses assume that he is faulty.

As we show in \cref{sec:eval}, the flat \coinname structure
can still quickly sign keyblocks for the day-long window (144 witnesses) 
while maintaining a throughput higher than Bitcoin currently supports.
Flat \coinname is more robust to faults, but increases the communication latency back
to $O(n)$.
Furthermore, if all faults ($\lfloor{\frac{N}{3}}\rfloor$ ) are consecutive leaders,
this can lead back to a worst case $O(n^2)$ communication latency.
%
%
%
%
%

\subsubsection{Membership Churn and BFT}\label{sec:3f+2}

After a new leader is elected, the system needs to ensure that 
the first microblock of the new leader points to the last microblock of the previous leader. 
Having $2f+1$ supporting votes is not enough. 
This occurs because there is the possibility 
than an honest node lost its membership when the new era started.
Now in the worst case, the system has $f$ Byzantine nodes,
$f$ honest nodes that are up to date, $f$ slow nodes that have a stale view of the blockchain, 
and the new leader that might also have a stale view. 
This can lead to the leader proposing a new microblock, 
ignoring some signed microblocks and getting $2f+1$ support (stale+Byzantine+his own). 
For this reason, the first microblock of an era needs $2f+2$ supporting signatures. 
If the leader is unable to obtain them, this means that he needs to synchronize with the system,
\ie, he needs to find the latest signed microblock from the previous roster. 
He asks all the nodes in his roster, plus the node that lost its membership,
to sign a latest-checkpoint message containing the hash of the last microblock.
At this point in time, the system has $3f+2$ ($3f+1$ of the previous roster plus the leader) members 
and needs $2f+1$ honest nodes to verify the checkpoint, 
plus an honest leader to accept it (a Byzantine leader will be the $f+1$ fault and compromise liveness). 
Thus, \coinname{} can tolerate $f$ fails in a total of $3f+2$ nodes.

\section{Performance Evaluation}\label{sec:eval}

In this section we discuss the evaluation of the \coinname prototype and our
experimental setup. The main question we want to evaluate is whether \coinname
is usable in practice without incurring large overheads. In particular we focus
on consensus latency and transaction throughput for different parameter
combinations.

\subsection{Prototype Implementation}\label{subsec:impl}

We implemented \coinname in Go\footnote{\url{https://golang.org}}
and made it publicly available on
GitHub.\footnote{\url{https://github.com/dedis/cothority}}  \coinname's
consensus mechanism is based on the CoSi protocol with Ed25519
signatures~\cite{bernstein12highspeed} and implements both flat- and tree-based
collective signing layouts as described in~\cref{sec:arch}. For comparison, we
also implemented a conventional PBFT consensus algorithm with the same
communication patterns as above and a consensus algorithm that uses individual
signatures and tree-based communication.

To simulate consensus groups of up to
$1008$ nodes, we oversubscribed the available $36$ physical machines (see below) and
ran up to $28$ separate \coinname processes on each server. Realistic wide-area
network conditions are mimicked by imposing a round-trip latency of $200$\,ms
between any two machines and a link bandwidth of $35$\,Mbps per simulated host.
Note that this simulates only the connections between miners of the consensus
group and not the full Bitcoin network. Full nodes and clients are not part of
the consensus process and can retrieve signed blocks only after consensus is
reached. Since Bitcoin currently is rather centralized and has only a few dozen
mining pools~\cite{apostolaki16hijacking}, we assume that if/when
decentralization happens, all miners will be able to support these rather
constrained network requirements. 

The experimental data to form microblocks was
taken by \coinname clients from the actual Bitcoin blockchain.  Both micro- and
keyblocks are fully transmitted and collectively signed through the tree and are
returned to the clients upon request together with the proof.  Verification of
block headers is implemented but transaction verification is only emulated to
avoid further measurement distortion through oversubscribed servers. A similar
practice is used in Shadow Bitcoin~\cite{miller15shadow}. We base our emulation
both on measurements~\cite{gervais15tampering} of the average block-verification
delays (around $200$\,ms for $500$\,MB blocks) and on the claims of Bitcoin
developers~\cite{scaling} that as far as hardware is concerned Bitcoin can
easily verify $4000$\,TPS. We simulate a linear increase of this delay
proportional to the number of transactions included in the block.
Because of the communication pattern of \coinname, the transaction-verification
cost delays only the leaf nodes.  By the time the leaf nodes finish the block
verification and send their vote back to their parents, all other tree nodes
should have already finished the verification and can immediately proceed.  For
this reason the primary delay factor is the transmission of the blocks that
needs to be done $\log N$ sequential times.

We ran all our experiments on DeterLab~\cite{Deterlab} using $36$ physical
machines, each having four Intel E5-2420 v2 CPUs and $24$\,GB RAM and being
arranged in a star-shaped virtual topology. 

\com{Finally the Bitcoin-core developers state~\cite{scaling} 
that as far as transaction verification is concerned,
Bitcoin nodes can handle up to 4000 TPS, 
hence the less than 1000 TPS that \coinname{} achieves 
will not be affected by the transaction verification delay.}
\com{
\paragraph{CoSi Pipelining.} Collective signing~\cite{syta15decentralizing} 
is done in four different phases per round, 
namely announce, response, challenge, and commit.
In \coinname the announce and commit phases of CoSi
can be performed in advance before the block to be committed is available,
since the proposed block can be sent to the signers in the challenge phase. 
This enables us to pipeline two rounds 
so that the announce/commit phases of
\coinname's commit round are piggybacked
on the challenge and response messages of the prepare round. 
This pipelining reduces latency by one round-trip over the communication tree.
Looking into the normal execution of \coinname, this pipeline 
can be extended so that an optimal throughput 
of one signed microblock per round-trip is produced. 
A sample execution can be seen in \cref{tab:byzcoin-pipeline}.


\begin{table}[t]
    \centering
    \scriptsize
    \begin{tabular}{llccccc}
        \toprule
                                   &            & $t_i$ & $t_{i+1}$ & $t_{i+2}$ & $t_{i+3}$ & $t_{i+4}$ \\
        \midrule
        \multirow{2}{*}{$B_k$}     &  prepare   & An/Co & Ch/Re & & & \\
                                   &  commit    & & An/Co & Ch/Re & & \\
        \midrule
        \multirow{2}{*}{$B_{k+1}$} &  prepare   & & An/Co & Ch/Re & & \\
                                   &  commit    & & & An/Co & Ch/Re & \\
        \midrule
        \multirow{2}{*}{$B_{k+2}$} &  prepare   & & & An/Co & Ch/Re & \\
                                   &  commit    & & & & An/Co & Ch/Re \\
        \bottomrule
    \end{tabular}
    \caption{\coinname pipelining for maximum transaction-throughput; $B_k$
    denotes the microblock signed in round $k$, An/Co the Announce-Commit and
    Ch/Re the Challenge-Response round-trips of CoSi}
    \label{tab:byzcoin-pipeline}
\end{table}
}

\com{
\subsection{Evaluation}\label{subsec:eval} 
The main question we want to evaluate is
whether \coinname{} is usable in practice without incurring large overhead. 
We evaluated keyblock and microblock signing
for an increasing number of consensus group members.
Further, we evaluated
\coinname's latency for an increasing microblock size and an increasing number of consensus group members,
for all implemented consensus algorithms.
We then compare Bitcoin with the flat and tree-based versions of \coinname 
to investigate the maximum throughput that each variant can achieve. 
In another experiment, we investigate \com{the impact of link bandwidth on \coinname
and} the latency of signing single transactions and larger blocks. 
Finally, we demonstrate the practicality of \coinname,
as far as latency for securing transactions is concerned, 
from a client's point of view.
}


\subsection{Consensus Latency}

The first two experiments focus on how microblock commitment latency
scales with consensus group size
and with number of transactions per block.

\subsubsection{Consensus Group Size Comparison}

This experiment focuses on the scalability of \coinname's BFT protocol
in terms of the consensus group size.
The number of unique miners participating in a consensus group
is limited by the number of membership shares in the window
(\cref{sec:membership}),
but can be smaller if some miners hold multiple shares
(\ie, successfully mined several blocks) within the same window.

We ran experiments for Bitcoin's 
maximum block size (1\,MB) with a variable number of participating hosts. 
Every time we increased the number of hosts,
we also increased the servers' bandwidth so that the available bandwidth
per simulated host remained constant (35\,Mbps). For the PBFT simulation, the 1\,MB block was too big to handle, thus the PBFT line corresponds to a 250\,KB block size.

\diagram{Influence of the consensus group size on the consensus latency}{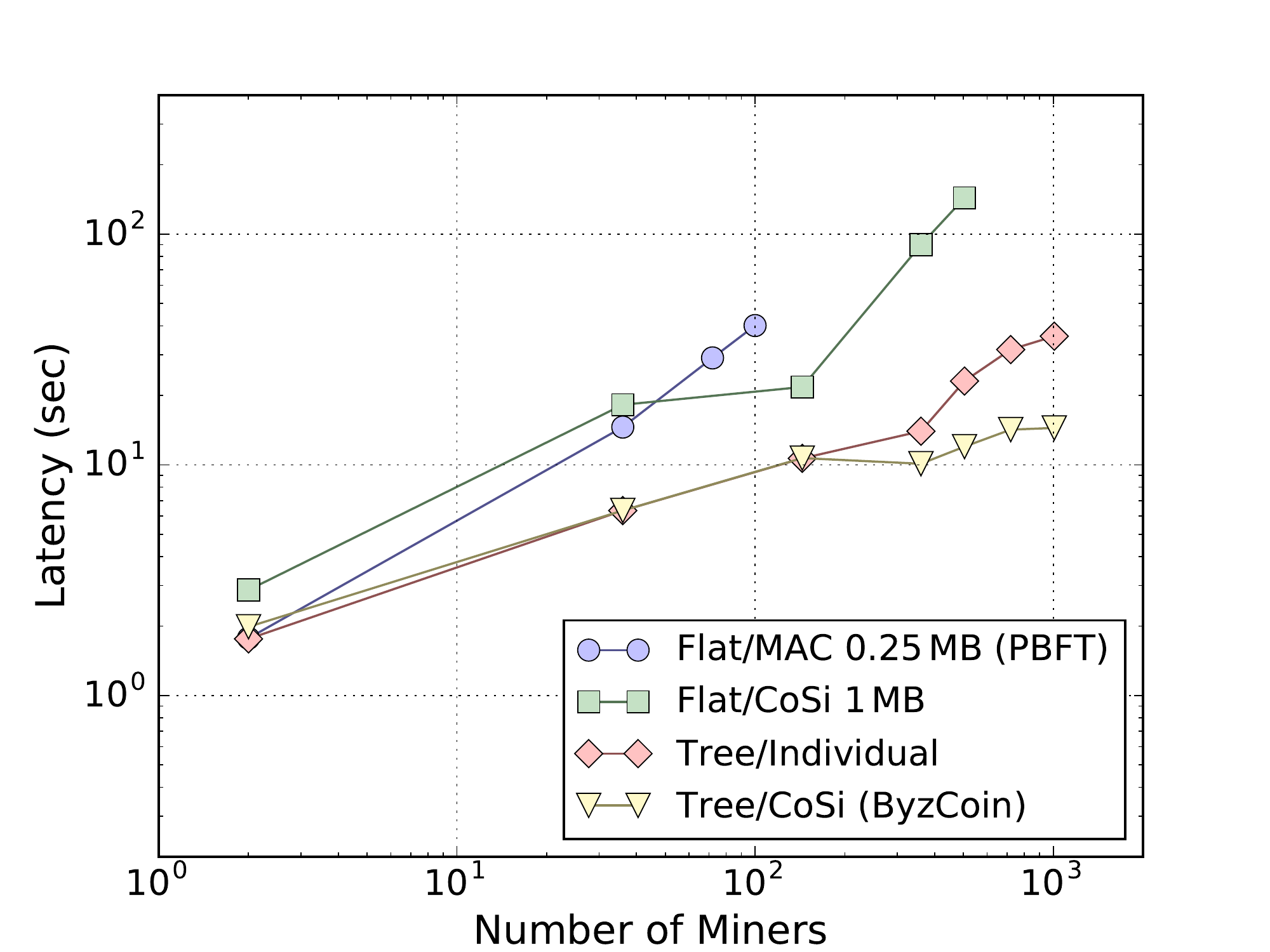}

As \cref{fig:scale_host.pdf} shows, the simple version of
\coinname achieves acceptable latency, 
as long as the consensus group size is less than 200. 
After this point the cost for the leader to broadcast the block to everyone 
incurs large overheads.
On the contrary, the tree-based \coinname scales well, 
since the same $1$\,MB block for $1008$ nodes suffers signing latency less
than the flat approach for $36$ nodes.
Adding $28$ times more nodes (from $36$ to $1008$) 
causes a latency increase close to a factor $2$ (from $6.5$ to $14$ seconds).
The basic PBFT implementation is quite fast for $2$ nodes 
but scales poorly ($40$ seconds for $100$ nodes),
whereas the tree-based implementation with individual signatures 
performs the same as \coinname for up to $200$ hosts.
If we aim for the higher security level of $1008$ nodes, however,
then \coinname is $3$ times faster.

\cref{fig:key_block_signing.pdf} shows the performance cost of keyblock signing.
The flat variant outperforms the tree-based version when the number of hosts is
small since the blocks have as many transactions as there are hosts and thus are
small themselves. This leads to a fast transmission even in the flat case and
the main overhead comes from the block propagation latency, which scales with
$O(\log N)$ in the tree-based \coinname variant.

\diagram{Keyblock signing latency}{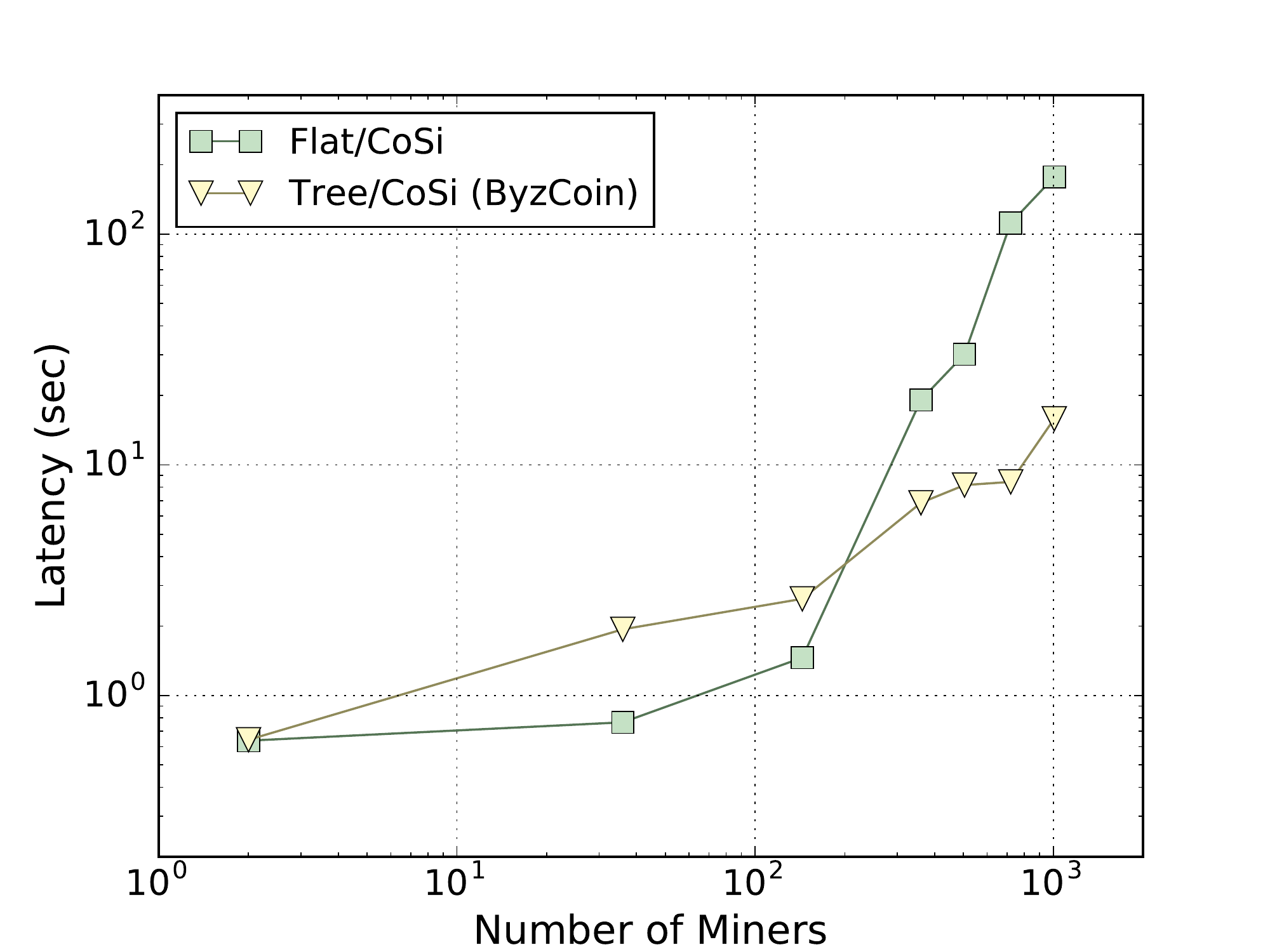}


\subsubsection{Block Size Comparison}

The next experiment analyzes how different block sizes affect the scalability of
\coinname. We used a constant number of $144$ hosts for all implementations.
Once again, PBFT was unable to achieve acceptable latency with $144$ nodes, 
thus we ran it with $100$ nodes only.

\cref{fig:scale_block.pdf} shows the average latency of the consensus mechanism, determined 
over $10$ blocks when their respective sizes increase.
As in the previous section we see that the flat implementation is
acceptable for a $1$\,MB block, 
but when the block increases to $2$\,MB the latency quadruples. 
This outcome is expected as the leader's link saturates 
when he tries to send $2$\,MB messages to every participating node. 
In contrast \coinname scales well because
the leader outsources the transmission of the blocks to other nodes and 
contacts only his children. 
The same behavior is observed for the algorithm that uses individual signatures and
tree-based communication, 
which shows that the block size has no negative effect on scalability when a tree is used.
Finally, we find that PBFT is fast for small blocks, 
but the latency rapidly increases to $40$ seconds for $250$\,KB blocks. 

\diagram{Influence of the block size on the consensus latency}{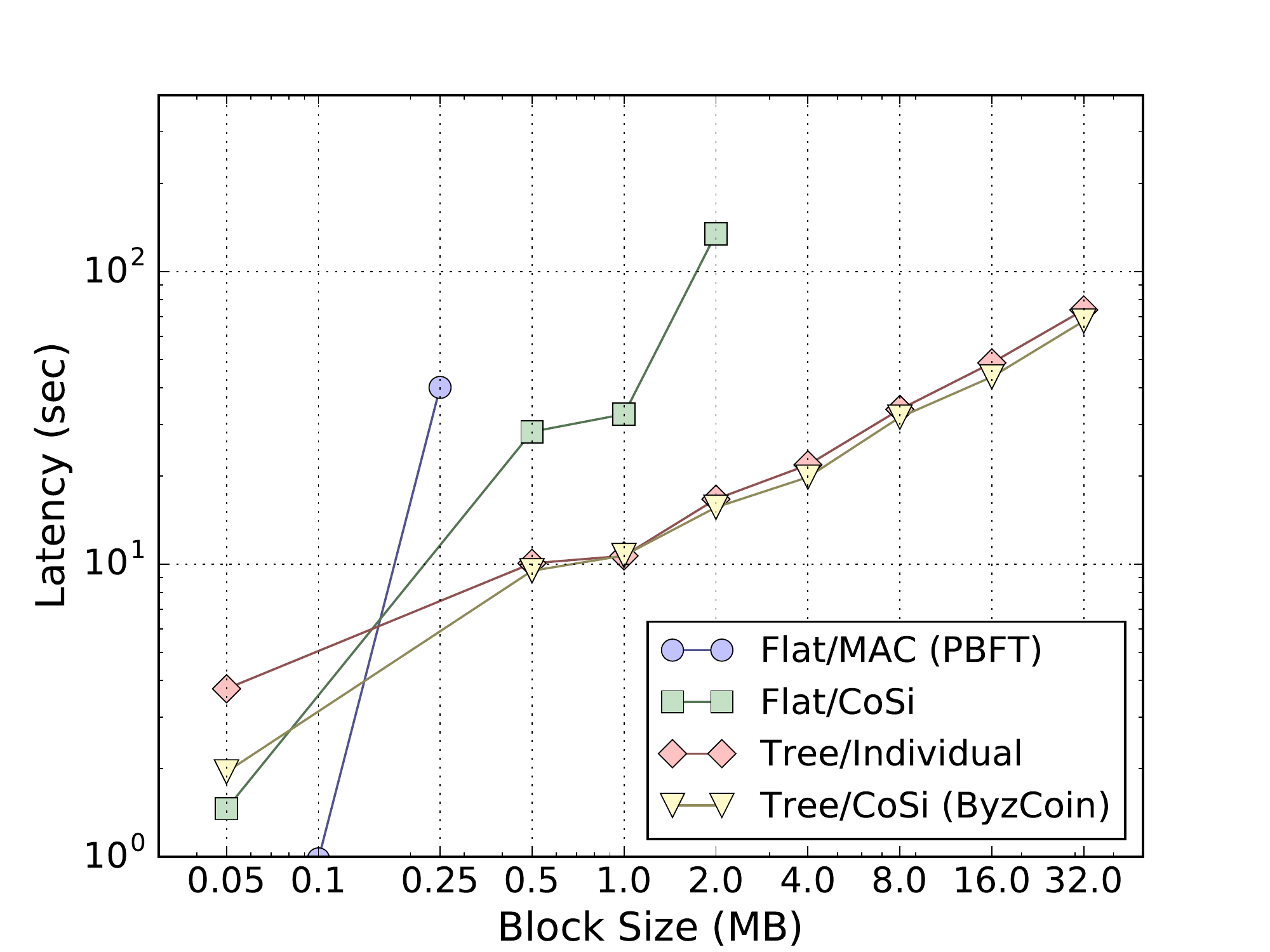}

\coinname's signing latency for a $1$\,MB block is close to
$10$ seconds, which should be
small enough to make the need for 0-confirmation transactions almost disappear. 
Even for a $32$\,MB block ($\approx 66000$ transactions)
the delay is much lower (around $90$ seconds) than the $\approx 10$ minutes required by Bitcoin.


\com{ I don't understand the paragraph below,
	and should we really be proposing new design extensions
	on an ad hoc basis deep in the evaluation?  Delete?  -baf 
For those cases where it is crucial that transactions are nevertheless completed
very quickly (\eg, less than $1$ minute for $32$\,MB blocks) we propose the
following defense against 0-confirmation attacks: All \coinname miners implement
a collective signing-based mechanism that allows them to process single
transactions and to issue a commitment to add those to the block after the next.
Adding them directly to the next block will interrupt 
and alter the mining process and also introduces race conditions, 
as a transaction can be signed while the block is being signed. 
This is also a corner-case where only $f$ faults out of $3f+2$
nodes can be tolerated.}

\cref{fig:block.pdf} demonstrates the signing latency of various blocks sizes on tree-based \coinname. 
Signing one-transaction blocks takes only $3$ seconds even when $1008$ miners co-sign it. 
For bigger blocks, we have included 
Bitcoin's current maximum block size of $1$\,MB
along with the proposed limits
of $2$\,MB in Bitcoin Classic
and $8$\,MB in Bitcoin Unlimited~\cite{andresen16classic}.
As the graph shows,
$1$\,MB and $2$\,MB blocks scale linearly in number of nodes at first
but after $200$ nodes, the propagation latency is higher 
than the transmission of the block, hence the latency is close to constant. 
For $8$\,MB blocks, even with $1008$ the signing latency increases only linearly.

\diagram{Influence of the consensus group size on the block signing latency}{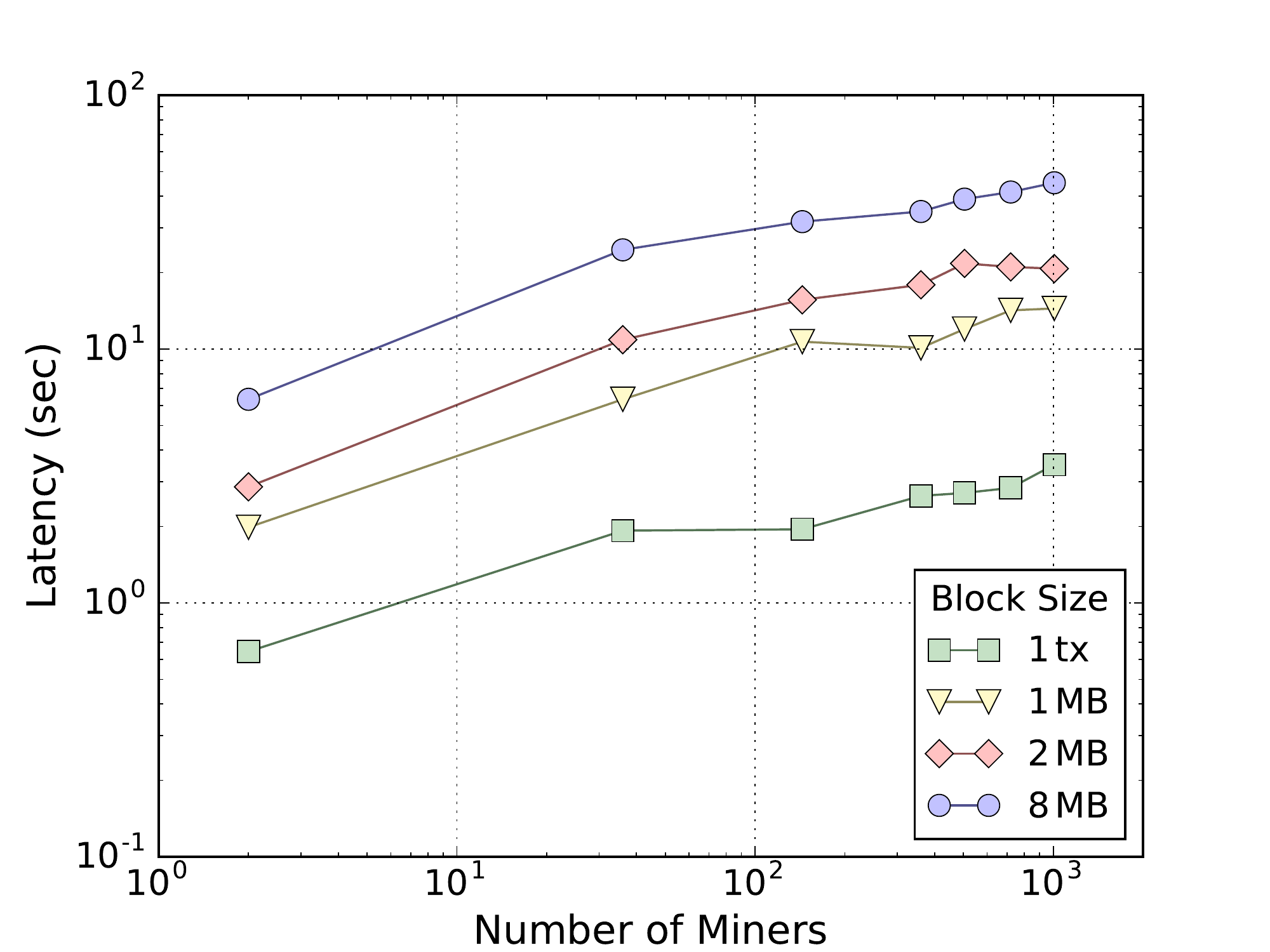}

\subsection{Transaction Throughput} 

In this experiment, we investigate the maximum throughput
in terms of transactions per second (TPS) that
\coinname can achieve, and show how Bitcoin could improve its throughput by
adopting a \coinname-like deployment model. We tested \coinname versions with
consensus group sizes of $144$ and $1008$ nodes, respectively. Note that
performance-wise this resembles the worst case scenario since the miner-share
ratio is usually not 1:1 as miners in the consensus group are allowed to hold
multiple shares, as described in~\cref{sec:membership}. 

Analyzing~\cref{fig:throughput.pdf} shows that Bitcoin can increase its overall
throughput by more than one order of magnitude through adoption of a flat
\coinname-like model, which separates transaction verification and block mining
and deals with forks via strong consistency. Furthermore, the $144$ node
configuration can achieve close to $1000$\,TPS,
which is double the throughput of
Paypal, and even the $1008$-node roster achieves close to $700$\,TPS. 
Even when the tree fails, the system can revert back to $1$\,MB microblocks on
the flat and more robust variant of \coinname and still have a throughput ten
times higher than the current maximum throughput of Bitcoin.

\diagram{Throughput of \coinname \com{Remove "Throughput Comparison" header
from picture, if possible increase picture size so that numbers on top of bars are not
outside of the frame;``ByzCoin'' $=>$ ``ByzCoin''; remove whitespace before
``Flat'' and write ``Flat-ByzCoin'' and not ``Flat -- ByzCoin'' }}{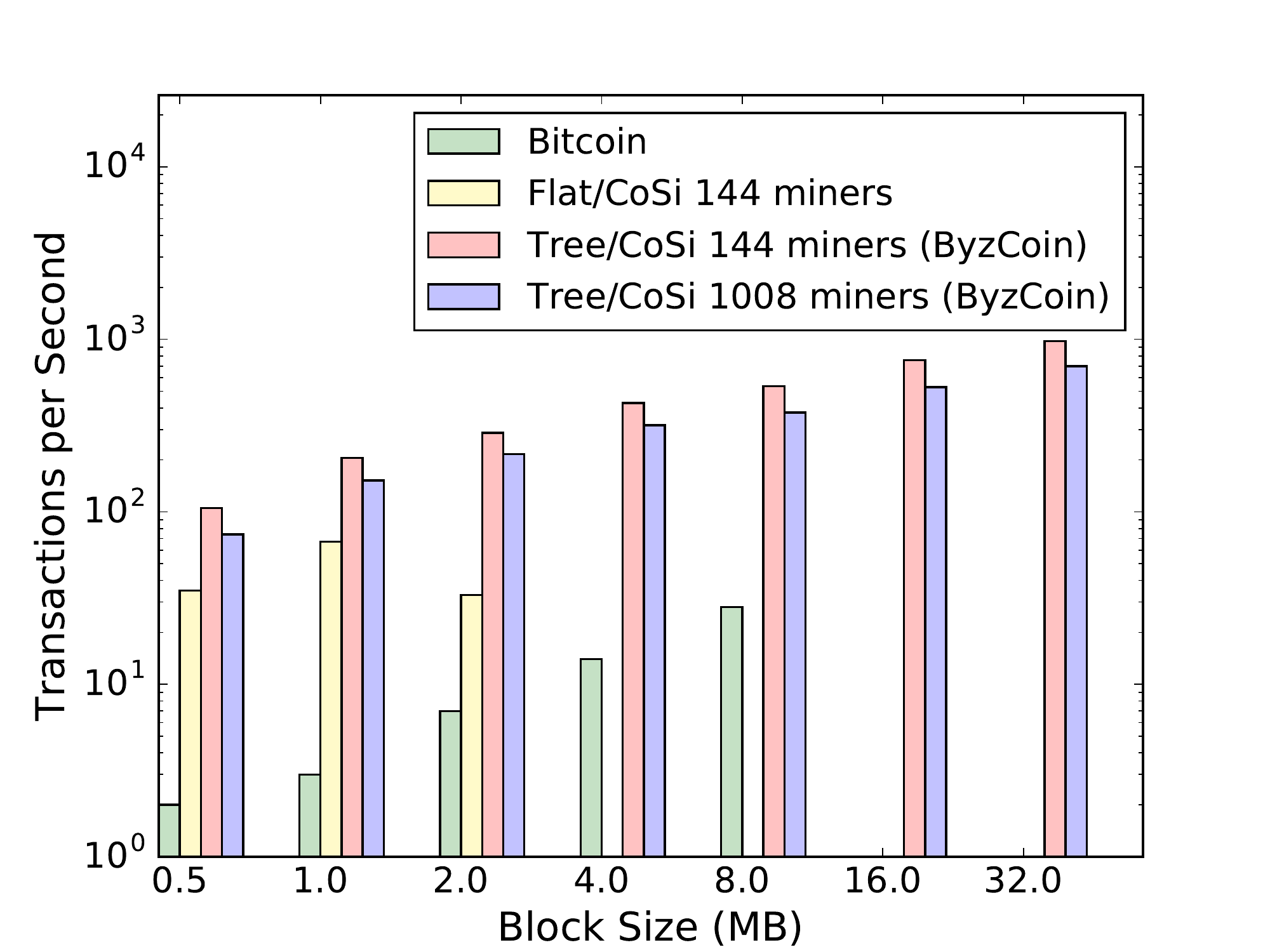}

\com{ The first several figures look good,
	but this figure (9, throughput) and the next two 
	double-spend and client-perceived latency) should be
	vertically compressed a lot more (flatter).
	Especially figs 10 and 11.
}

In both \cref{fig:scale_block.pdf,fig:throughput.pdf},
the usual trade-off between throughput and latency
appears. The system can work with
$1$--$2$\,MB microblocks when the load is normal and then has a latency of $10$--$20$ seconds.
If an overload occurs,
the system adaptively changes the block size to enable higher throughput.
We note that this is not the case in the simple \coinname
where $1$\,MB microblocks have optimal throughput and acceptable latency.

\com{\subsubsection{Bandwidth Changes}

In this experiment we tested how the tree-based \coinname variant behaves 
when changing link bandwidth.
As we see in \ref{fig:BW.eps} there is a knee point 
for each block size 
where any additional bandwidth increase has little effect 
on the system's performance. 
However, if the available bandwidth is lower than this knee point,
then performance degrades significantly. Thus, any system 
deploying \coinname should adapt the microblock size 
so that the available bandwidth is
higher than the knee point of the block size. 
If the load of the system increases the block size 
needs to increase too as \cref{fig:throughput.pdf} illustrates, 
which can only be of help if the bandwidth is higher than the knee point of the block size. 

\diagram{Latency versus link bandwidth \com{Link Bandwidth per Host (Mbps)}}{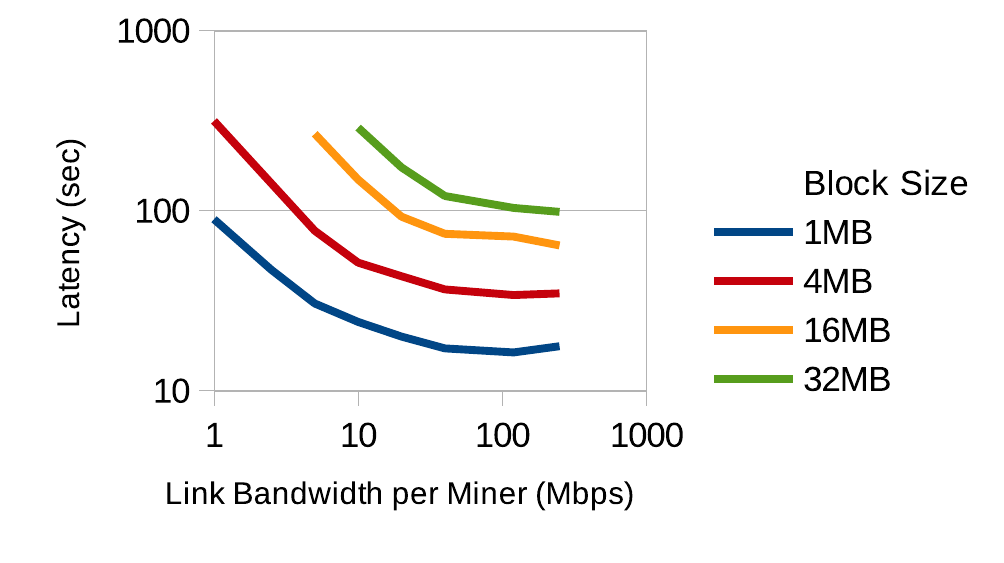}}

\section{Security Analysis}\label{sec:security}

In this section, we conduct a preliminary, informal security analysis of \coinname, and
discuss how its consensus mechanism can mitigate or eliminate some known
attacks against Bitcoin.

\com{ Do a more careful security analysis focused on ByzCoin,
	rather than focused on Bitcoin.  In particular,
	discuss the advantages and risks of the window-membership mechanism,
	e.g., risk of unavailability if too many members disappear;
	tradeoffs between large and small windows; etc.
	Risks of performance-DoS similar to other BFT implementations.
	etc.}

\subsection{Transaction Safety}

%

In the original Bitcoin paper~\cite{nakamoto08bitcoin},
Nakamoto models Bitcoin's security
against transaction double spending attacks
as in a Gambler's Ruin Problem. 
Furthermore, he models the progress an attacker can make 
as a Poisson distribution and combines these two models 
to reach~\cref{eq:secure}. 
This equation calculates the probability of a successful double spend
after $z$ blocks when the adversary controls $q$ computing power.

\begin{equation} \label{eq:secure} 
P = 1-\sum_{k=0}^{z} \frac{\lambda^{k} e^{-\lambda}}{k! } 
\left( 1 - \left( \frac{q}{p} \right) ^{ 
\left( z-k \right)} \right)
\end{equation}

In \cref{fig:attackers.pdf,fig:block_safety_latency.pdf} we compare the relative
safety of a transaction over time in Bitcoin\footnote{Based on
data from \url{https://blockchain.info}.} versus \coinname. \cref{fig:attackers.pdf} shows that \coinname
can secure a transaction in less than a minute,
because the collective signature guarantees forward security.
On the contrary, Bitcoin's transactions need hours 
to be considered fully secured from a double-spending attempt.
\cref{fig:block_safety_latency.pdf} illustrates the required time 
from transaction creation to the point where a double spending attack has less
than $0.1$\% chance of success.
\coinname incurs a latency of below one minute to achieve the above security,
which boils down to the time the systems needs to produce a collectively
signed microblock. Bitcoin on the other hand needs several hours to reach the
same guarantees.
Note that this graph does not consider other advanced attacks, such as 
eclipse attacks~\cite{heilman15eclipse}, 
where Bitcoin offers no security for the victim's transactions.

\diagram{Successful double-spending attack probability \com{Move
        legend outside of the picture; Probability
        for Successful Double-Spending ($\%$)
        Time Since Transaction Creation (min); Bitcoin-Attacker controls xx $=>$
Attacker controls xx}}{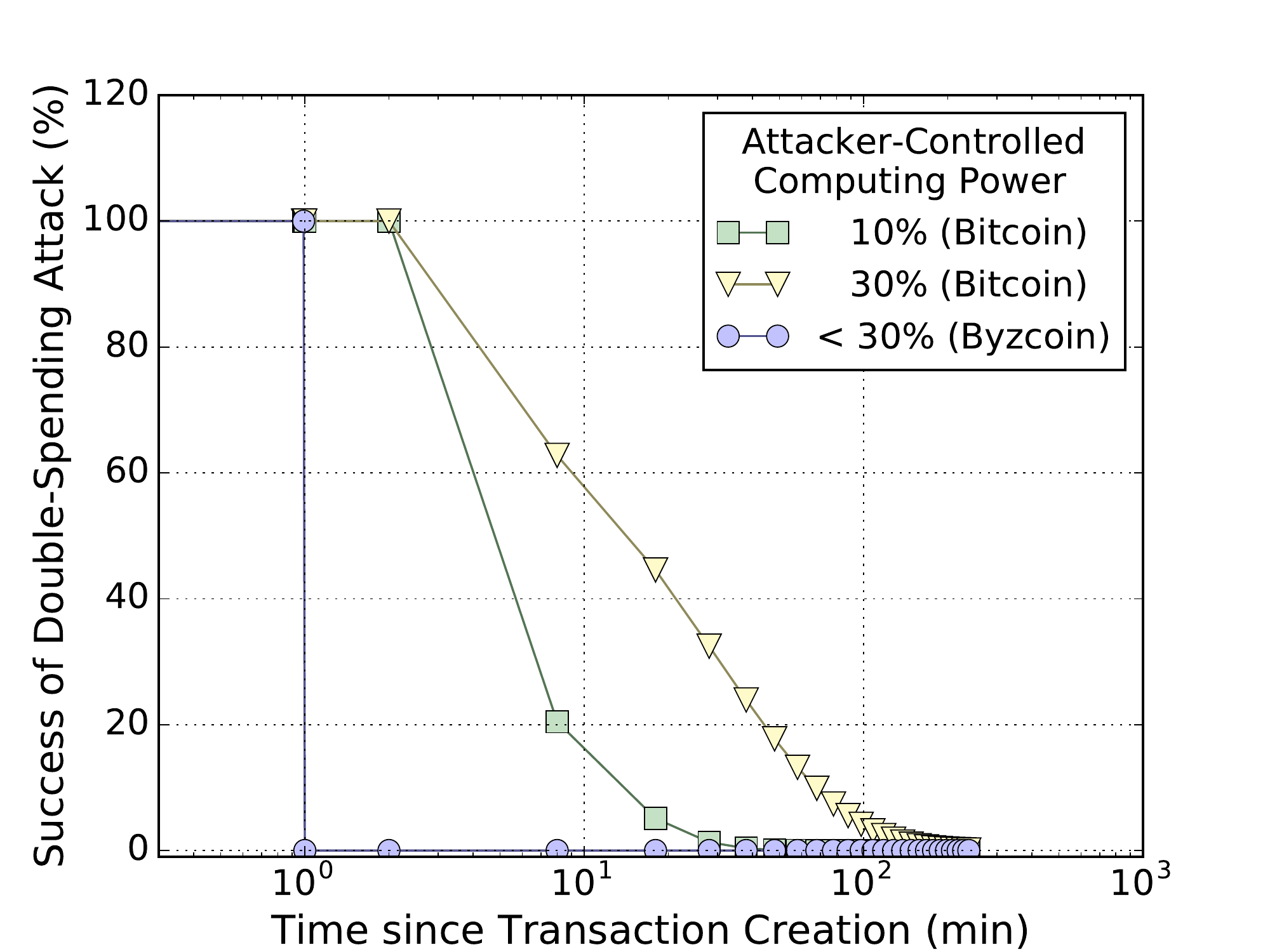}

\diagram{Client-perceived secure transaction latency
\com{BitCoSi $=>$ ByzCoin; Latency for Securing a
    Transaction (min); Attacker-controlled Computing Power ($\%$)}}{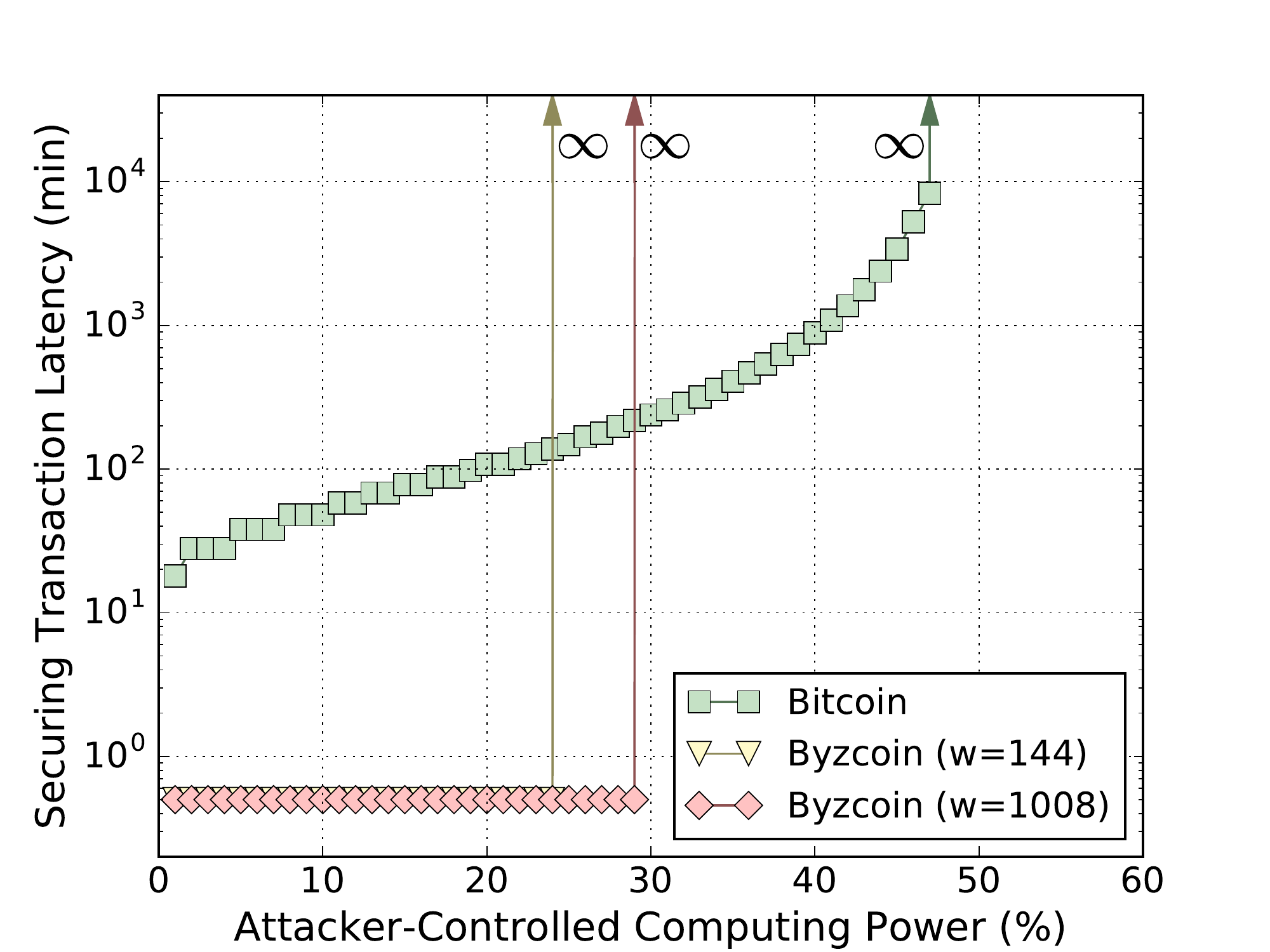}

%

\subsection{Proof-of-Membership Security}

The security of \coinname's proof-of-membership mechanism can be modeled as a
random sampling problem with two possible independent outcomes (honest,
Byzantine). The probability of picking a Byzantine node (in the worst case) is
$p = 0.25$ and the number of tries corresponds to the share window size $w$. In
this setting, we are interested in the probability that the system picks less
than $c = \lfloor{\frac{w}{3}}\rfloor$ Byzantine nodes as consensus group
members and hence guarantees safety. To calculate this probability, we use the
cumulative binomial distribution where $X$ is the random variable that
represents the number of times we pick a Byzantine node:

\begin{equation}
\label{eq:binom}
    P\left[X \leq c \right] = \sum_{k=0}^{c} {w \choose k} p^k \left(1-p\right)^{w-k}
\end{equation}

\cref{tab:safety} displays the results for the evaluation of~\cref{eq:binom} for
various window sizes $w$ both in the common threat model where an adversary
controls up to $25\%$ hash power and in the situation where the system faces
a stronger adversary with up to $30\%$ computing power. The latter might
temporarily occur due to hash power variations and resource churn.

\begin{table}[ht]
\centering
\caption{Expected proof-of-membership security levels}
\label{tab:safety}
\small
\begin{tabular}{rrrrrrr}
\toprule
$p~\vert~w$  & 12    & 100   & 144   & 288   & 1008  & 2016  \\
\midrule
0.25 & 0.842 & 0.972 & 0.990 & 0.999 & 0.999 & 1.000 \\
0.30 & 0.723 & 0.779 & 0.832 & 0.902 & 0.989 & 0.999 \\
\bottomrule
\end{tabular}
\end{table}

At this point, recall that $w$ specifies the number of available shares and not
necessarily the number of actual miners as each member of the consensus group
is allowed to hold multiple shares. This means that the number of available
shares gives an upper bound on the latency of the consensus mechanism with the
worst case being that each member holds exactly one share.

In order to choose a value for $w$ appropriately one must take into account
not only consensus latency and the desired security level (ideally $\geq 99\%$)
but also the increased chance for resource churn when values of $w$ become
large. From a security perspective the results of~\cref{tab:safety} suggest that
the share window size should not be set to values lower than $w = 144$. Ideally,
values of $w = 288$ and above should be chosen to obtain a reasonable security
margin and, as demonstrated in~\cref{sec:eval}, values up to $w = 1008$ provide
excellent performance numbers.

Finally, care should be taken when bootstrapping the protocol, as for small
values of $w$ there is a high probability that a malicious adversary is able to
take over control. For this reason we suggest that \coinname starts with vanilla
Nakamoto consensus and only after $w$ keyblocks are mined the \coinname
consensus is activated.

\com{TODO: is it possible to use bias-resistant public randomness to pick the
first $w$ members for bootstrapping which might allow to avoid a long initial
delay before \coinname consensus can be used?}

\subsection{Defense Against Bitcoin Attacks}

\paragraph{0-confirmation Double-Spend Attacks.}

Race~\cite{karame12double} and Finney~\cite{Finney} attacks belong to the family
of 0-confirmation double-spend attacks which might affect traders that provide
real-time services to their clients. In such scenarios the time between exchange
of currency and goods is usually short because traders often cannot afford to wait
an extended period of time (10 or more minutes) until a transaction they
received can be considered indeed confirmed.

\coinname can mitigate both attacks by putting the merchant's transaction in a
collectively signed microblock whose verification latency is in the order of a
few seconds up to a minute. If this latency is also unacceptable, then he can
send a single transaction for signing, which will cost more, but is secured in
less than $4$ seconds. 

\paragraph{N-confirmation Double-Spend Attacks.} 

The assumption underlying this family of attacks~\cite{NConf} is that, after
receiving a transaction for a trade, a merchant waits $N-1$ additional
blocks until he concludes the interaction with his client. At this point, a
malicious client creates a new double-spending transaction and tries to fork the
blockchain, which has a non-negligible success-probability if the adversary has
enough hash power. For example, if $N=3$ then an adversary holding $10\%$
of the network's hash power has a $5\%$ success-chance to mount the above
attack~\cite{nakamoto08bitcoin}.

In \coinname the merchant would simply check the collective signature of the
microblock that includes the transaction, which allows him to verify that it was accepted
by a super-majority of the network. Afterwards the attacker cannot succeed in
forking the blockchain as the rest of the signers will not accept his new block.
Even if the attacker is the leader, the proposed microblock will be rejected,
and a view change will occur.

\paragraph{Eclipse and Delivery-Tampering Attacks.}

In an eclipse attack~\cite{heilman15eclipse} it is assumed that an adversary
controls a sufficiently large number of connections between the victim and the
Bitcoin network. This enables the attacker to mount attacks such as 0- and
N-confirmation double-spends with an ever increasing chance of success the
longer the adversary manages to keep his control over the network.
Delivery-tampering attacks~\cite{gervais15tampering} exploit Bitcoin's
scalability measures to delay propagation of blocks without causing a network
partition. This allows an adversary to control information that the victim
receives and simplifies to mount 0- and 1-confirmation double-spend attacks as
well as selfish-mining.

While \coinname does not prevent an attacker from eclipsing a victim
or delaying messages in the peer-to-peer network,
its use of collective signatures in transaction commitment
ensure that a victim cannot be tricked into accepting an
alternate attacker-controlled transaction history
produced in a partitioned network fragment.

\paragraph{Selfish and Stubborn Mining Attacks.}

Selfish mining~\cite{eyal14majority} allows a miner to increase his profit by
adding newly mined blocks to a hidden blockchain instead of instantly
broadcasting them. This effect can be further amplified if the malicious miner
has good connectivity to the Bitcoin network. The authors of selfish mining
propose a countermeasure that thwarts the attack if a miner has less than $25\%$
hash power under normal circumstances or less than $33\%$ in case of an
optimal network. Stubborn mining~\cite{nayak15stubborn} further generalizes the
ideas behind selfish mining and combines it with eclipse attacks in order to
increase the adversary's revenue.

In \coinname, these strategies are ineffective as forks are instantly resolved in a
deterministic manner, hence building a hidden blockchain only wastes resources
and minimizes revenue. Another approach to prevent the above attacks would be to
include bias-resistant public randomness~\cite{lenstra15random} in every
keyblock. This way even if an attacker gains control over the consensus
mechanism (\eg, by having $>33\%$ hash power) he would still be unable to
mine hidden blocks. We leave exploring this approach for future research.

\paragraph{Transaction Censorship.}

In Bitcoin-NG, a malicious leader can censor transactions for the duration of his epoch(s). 
The same applies for \coinname. 
However, as not every leader is malicious, the censored transactions are
only delayed and will be processed eventually by the next honest
leader. 
\coinname can improve on this, as the leader's actions are double-checked by all the other
miners in the consensus group.
A client can announce his censored transaction
just like in classic PBFT; this will indicate a potential leader fault and will
either stop censorship efforts or lead to a view-change to remove
\com{
the malicious leader. Finally, using fork-based attacks to censor transactions,
\eg, conducted by adversaries who control significant amounts of the overall
hash power, is no longer possible due to \coinname's deterministic fork
}%
the malicious leader. Finally, in Bitcoin(-NG) a miner can announce his intention
to fork over a block that includes a transaction, 
giving an incentive to other miners to exclude this transaction.
In \coinname using fork-based attacks to censor transactions 
is no longer possible due to \coinname's deterministic fork
resolution mechanism. An attacker can therefore only vote down a leader's
proposals by refusing to co-sign. This is also a limitation, however, 
as an adversary who controls more than 33\% of the shares (\cref{sec:future})
deny service and can censor transactions for as long as he wants.

%

\section{Related Work}\label{sec:related}

\com{ Other BFT-related Bitcoin papers we forgot to cite
	and probably should:

	Tendermint
	\url{http://tendermint.com/docs/tendermint.pdf}.
	Proposes using BFT but for closed-group, "permissioned" ledgers.
	Proof-of-stake mechanism.  Discusses transaction censorship problem.

	HoneyBadger
	\url{http://eprint.iacr.org/2016/199.pdf}.
	Focuses on BFT in unpredictable-latency asynchronous settings.
	Good paper to cite in context of performance-slowdown attacks.

	Christian Cachin's asynchronous BFT work: e.g.,
	~\cite{cachin00random,cachin01secure,ramasamy06parsimonious},
	also uses threshold signatures as a primitive,
	though without tree-based aggregation.
}

\com{
How to handle Bitcoin's performance shortcomings is currently one of the major
controversies discussed in the community. But even though it is a urgent
concern, so far no solution was found that all involved parties were satisfied
with. A simple approach would be to increase the block size and with that the
maximum throughput of Bitcoin. However, this might also negatively affect the
security of the system and does not address the core of the problem: Bitcoin's
probabilistic consistency.}

\coinname and Bitcoin~\cite{nakamoto08bitcoin} share the same primary objective:
implement a state machine replication (SMR) system with open
membership~\cite{bonneau15research,garay15bitcoin}. Both therefore differ from
more classic approaches to Byzantine fault-tolerant SMRs with static or slowly
changing consensus groups such as PBFT~\cite{castro99practical},
Tendermint~\cite{buchman16tendermint}, or Hyperledger~\cite{hyperledger}. 

Bitcoin has well-known performance shortcomings; there are several
proposals~\cite{lewenberg15inclusive,wood14ethereum} on
how to address these. The GHOST
protocol~\cite{sompolinsky13fastbitcoin} changes the chain selection rule
when a fork occurs. While Bitcoin declares the fork with the most proof-of-work
as the new valid chain, GHOST instead chooses the entire subtree that received
the most computational effort. Put differently, the subtree that was considered
correct for the longest time will have a high possibility of being selected,
making an intentional fork much harder. One limitation of GHOST is that
no node will know the full tree, as invalid blocks are not propagated. 
While all blocks could be propagated, this makes the system
vulnerable to DoS attacks since an adversary can simply flood the network with
low-difficulty blocks.

Off-chain transactions, an idea that originated from the
two-point channel protocol~\cite{hearn15rapidly}, are another alternative to
improve latency and throughput of the Bitcoin network. Other similar proposals
include the Bitcoin Lightning Network~\cite{poon15bitcoin} and micro-payment
channels~\cite{decker15fast}, which allow transactions without a trusted
middleman. They use contracts so that any party can generate proof-of-fraud on
the main blockchain and thereby deny revenue to an attacker. Although these
systems enable faster cryptocurrencies, they do not address the core problem of
scaling SMR systems, thus sacrificing the open and distributed nature of
Bitcoin. Finally, the idea behind sidechains~\cite{back14enabling} is to
connect multiple chains with each other and enable the transfer of Bitcoins from
one chain to another. This enables the workload distribution to multiple subsets
of nodes that run the same protocol. 


There are several proposals that, like \coinname, target the consensus mechanism
and try to improve different aspects. Ripple~\cite{schwartz14ripple} implements
and runs a variant of PBFT that is low-latency and based on collectively-trusted
subnetworks with slow membership changes. The degree of decentralization
depends on the concrete configuration of an instance.
Tendermint~\cite{buchman16tendermint} also implements a PBFT-like
algorithm, but evaluates it with at most $64$ ``validators''.
Furthermore, Tendermint does not address important challenges
such as the link-bandwidth between validators,
which we found to be a primary bottleneck.
PeerCensus~\cite{decker16bitcoin} is an interesting alternative that shares
similarities with \coinname, but is only a preliminary theoretical analysis.

Finally, Stellar~\cite{mazieres15stellar} proposes a novel consensus protocol
named Federated Byzantine Agreement, which introduces Quorum slices that enable a
BFT protocol ``open for anyone to participate''. Its security, however, depends
on a nontrivial and unfamiliar trust model requiring
correct configuration of trustees by each client.

\section{Limitations and Future Work}\label{sec:future}

\com{This section still needs cleanup.}

This section briefly outlines several of \coinname's important
remaining limitations, and areas for future work.

\paragraph{Consensus-Group Exclusion.} A malicious \coinname leader can potentially exclude
nodes from the consensus process. This is easier in the flat variant, where the
leader is responsible for contacting every participating miner, but it is also
possible in the tree-based version, if the leader ``reorganizes'' the tree and
puts nodes targeted for exclusion in subtrees where the roots are colluding
nodes.  A malicious leader faces a dilemma, though: excluded nodes lose their
share of newly minted coins which increases the overall value per coin and thus
the leader's reward. The victims, however, will quickly broadcast view-change
messages in an attempt to remove the Byzantine leader.

As an additional countermeasure to mitigate such an attack, miners could run a
peer-to-peer network on top of the tree to communicate protocol details. Thus
each node potentially receives information from multiple sources. If the parent
of a node fails to deliver the announcement message of a new round, this node
could then choose to attach itself (together with its entire subtree) to another
participating (honest) miner. This self-adapting tree could mitigate the
leader's effort to exclude miners. As a last resort, the malicious leader could
exclude the commitments of the victims from the aggregate commitment, but as
parts of the tree have witnessed these commitments, the risk of triggering a
view-change is high. 

In summary, the above attack seems irrational as the drawbacks of trying to
exclude miners seem to outweigh the benefits. We leave a more thorough analysis
of this situation for future work.

\paragraph{Defenses Against 33\%+ Attacks.} An attacker powerful enough to
control more than $\frac{1}{3}$ of the consensus shares can, in the Byzantine
threat model, trivially censor transactions by withholding votes, and
double-spend by splitting honest nodes in two disjoint groups and collecting
enough signatures for two conflicting microblocks.
\cref{fig:block_safety_latency.pdf} shows how the safety of \coinname fails at
$30\%$, whereas Bitcoin remains safe even for $48\%$, if a client can wait long
enough.

However, the assumption that an attacker completely controls the network is
rather unrealistic, especially if messages are authenticated and spoofing is
impossible~\cite{apostolaki16hijacking}. The existence of the peer-to-peer
network on top of the tree, mentioned in the previous paragraph,
enables the detection of equivocation attacks such as microblock forks and
mitigates the double-spending efforts, as honest nodes will stop following the
leader. Thus, double-spending and history rewriting attacks in \coinname become
trivial only after the attacker has $66\%$ of the shares, effectively increasing
the threshold from $51\%$ to $66\%$. This assumption is realistic, as an
attacker controlling the complete network can actually split Bitcoin's network
in two halves and trivially double-spend on the weaker side. This is possible
because the weak side creates blocks that will be orphaned once the partition
heals.
We again leave a more thorough analysis of this situation for future work.

\com{Another drawback is the fact that an attacker or group of attackers that 
has more than 33\% of the computing power 
can censor transactions on someone by not signing their blocks with 
low cost. 
In Bitcoin censoring is possible by forking over any unwanted blocks, 
but if the attempt fails the miners lose their revenue. 
In \coinname not voting a block as valid only has the consequence of 
losing the transaction fees, 
which at this point is a lot less than the reward of mining a block. 
However censorship is part the orthogonal problem of privacy which is 
out of the scope of this paper.}

\com{Bryan's mail
Yes, they can act maliciously at >33\% threshold, but the question is what exactly they can accomplish at that threshold in practice.

I think perhaps the worst attack that becomes “easy” at that threshold is simply denial-of-service: i.e., the attacker can make everything grind to a halt by preventing a 66\% threshold from forming to prepare or commit transactions.  But in practice if that were to happen, in the worst case the angry admins of the rest of the system could forcibly remove a sufficient number the failed/misbehaving members from the consensus group so that the commit threshold becomes 66\% of the resulting smaller group and liveness is restored.  This would of course be bad and a pain for everyone, but it wouldn’t be remotely subtle and t doubt it would come to this in a real system unless there was some kind of catastrophic breakdown of cooperation incentives.

Double-spending and history-rewriting attacks, on the other hand, do not automatically become “easy” when the attacker has 34\%, because the attacker must still get approx 33\% worth of honest nodes to sign off on its commits.  The 33\% threshold actually comes from a worst-case security assumption where we assume that the attacker holds not only 33\% of the consensus group but also completely controls the network, e.g., can arbitrarily keep entire sets of honest nodes partitioned from each other.  Under that assumption, the attacker can use its 34\% and one set of 33\% of honest nodes to create one history, then (for example) double-spend using its 34\% and the other set of 33\% honest nodes to create the alternate history.  But the attacker must carefully ensure that those two big sets of honest nodes cannot or do not communicate.

If there’s any independent connectivity and regular peer-to-peer communication (e.g., gossip) between those honest nodes as well apart from the (malicious) leader’s actions, then any kind of double-spending or history-rewriting attack would probably be implausibly difficult in practice until the attacker achieves closer to a 66\% threshold.  In any deployment version of ByzCoin, we would definitely want that kind of peer-to-peer gossip as an additional security measure, even if it’s not necessary or even formally a benefit in terms of the theoretical worst-case security model where the adversary completely controls the net.  I also have some further provisions in mind to strengthen the system against exclusion attacks by a malicious leader (leaving non-failed nodes out of its prepare/commit trees), which would further strengthen the system against double-spending or equivocation attacks.

}

\paragraph{Proof-of-Work Alternatives.}
Bitcoin's hash-based proof-of-work has many drawbacks,
such as energy waste
and the efficiency advantages of custom ASICs
that have made mining by ``normal users'' impractical.
Many promising alternatives are available,
such as memory-intensive puzzles~\cite{ateniese14proofs}\com{Litecoin,
Argon2?}, or
proof-of-stake designs~\cite{king12ppcoin}.
Consensus group membership might in principle also be based on
other Sybil attack-resistant methods,
such as those based on social trust networks~\cite{yu08sybillimit}.
A more democratic alternative
might be to apportion mining power on a ``1 person, 1 vote'' principle,
based on anonymous {\em proof-of-personhood} tokens
distributed at pseudonym parties~\cite{ford08nyms}.
Regardless, we treat the ideal choice of Sybil attack-resistance mechanism
as an issue for future work, orthogonal to the focus of this paper.

\paragraph{Other Directions.} Besides the issues outlined above, there are many
more interesting open questions worth considering:
Sharding~\cite{croman16scaling} presents a promising approach
to scale distributed protocols and was already studied for private
blockchains~\cite{danezis16centrally}. A sharded variant of \coinname might thus
achieve even better scalability and performance numbers. A key obstacle that
needs to be analyzed in that context before though is the generation of
bias-resistant public randomness~\cite{lenstra15random} which would enable to
pick members of a shard in a distributed and secure manner. Yet another
challenge is to find ways to increase incentives of rational miners to remain
honest, like binding coins and destroying them when misbehavior is
detected~\cite{buchman16tendermint}. Finally, asynchronous
BFT~\cite{cachin00random,cachin01secure} is another interesting class of
protocols, which only recently started to be analyzed in the context of
blockchains~\cite{miller16honey}.

\section{Conclusion}\label{sec:conclusion}

\coinname is
a scalable Byzantine fault tolerant consensus algorithm
for open decentralized blockchain systems such as Bitcoin.
\coinname's strong consistency
increases Bitcoin's core security guarantees---shielding against attacks on the
consensus and mining system such as N-confirmation double-spending, intentional
blockchain forks, and selfish mining---and also enables high scalability and
low transaction latency. \coinname's application to Bitcoin is just one
example, though: theoretically, it can be deployed to any blockchain-based
system, and the proof-of-work-based leader election mechanism might be changed
to another approach such as proof-of-stake. If open membership is
not an objective, the consensus group could be static,
though still potentially large. We developed a wide-scale
prototype implementation of \coinname, validated its efficiency with
measurements and experiments, and have shown that Bitcoin can increase the
capacity of transactions it handles by more than two orders of magnitude.


\section*{Acknowledgments}
We would like to thank the DeterLab project team 
for providing the infrastructure for our
experimental evaluation, Joseph Bonneau
for his input on our preliminary design, and the anonymous 
reviewers for their helpful feedback.

\bibliographystyle{acm}
\bibliography{main}	

\end{document}